\documentclass[twocolumn,showpacs,preprintnumbers,amsmath,amssymb]{revtex4}

\usepackage{graphicx}
\usepackage{dcolumn}
\usepackage{bm}

\begin{document}


\title{Optical investigations of quantum-dot spin dynamics}


\author{Jan Dreiser}
\email{dreiser@phys.ethz.ch}
\author{Mete Atat\"ure}%
\author{Christophe Galland}
\author{Tina M\"uller}
\author{Antonio Badolato}
\author{Atac Imamoglu}
\affiliation{%
Institute of Quantum Electronics, ETH Z\"urich, Wolfgang-Pauli-Strasse, CH-8093 Z\"urich, Switzerland
}%

\date{\today}

\begin{abstract}
We have performed all-optical measurements of spin
relaxation in single self-assembled InAs/GaAs quantum dots (QD)
as a function of static external electric and magnetic fields.
To study QD spin dynamics  we measure the degree of resonant absorption which results from a competition between optical spin pumping induced by the resonant laser field and spin relaxation induced by reservoirs. Fundamental interactions that determine spin dynamics in QDs are
hyperfine coupling to QD nuclear spin ensembles,  spin-phonon
coupling  and exchange-type interactions with a nearby Fermi sea
of electrons. We show that the strength of spin relaxation 
generated by the three fundamental interactions can be changed by up to
five orders of magnitude upon varying the applied electric and
magnetic fields. We find that the strength of optical spin pumping
that we use to study the spin relaxation is determined
predominantly by hyperfine-induced mixing of single-electron spin
states at low magnetic fields and heavy-light hole mixing at high magnetic fields. Our measurements allow us to determine the rms value of the hyperfine
(Overhauser) field to be $\sim$15 mTesla with an electron g-factor of $g_e$=0.6 and a hole mixing strength of $|\varepsilon|^2=5\times 10^{-4}$.
\end{abstract}

\maketitle

\section{\label{sec:intro}introduction}

A single quantum dot (QD) electron spin is a fundamental physical system which allows for a controlled study of confined spin
dynamics in the solid-state. In contrast to higher-dimensional
semiconductor structures QD spins have been demonstrated to
posses long relaxation and coherence times exceeding 20 msec and
10 $\mu$sec respectively. These findings along with
demonstration of single spin manipulation and read-out have
strengthened the proposals for using QD spins as physical
representation of qubits in quantum information processing \cite{LossPRA98,ImamogluPRL99,CalarcoPRA03}. The prolongation of spin relaxation times for QD spins stems from a
drastic reduction in spin-phonon coupling mediated by a
combination of electron-phonon and spin-orbit interactions and
suppressed by strong quantum confinement of electrons. As a
consequence additional spin-reservoir interactions such as
hyperfine coupling to QD nuclear spins and exchange-type
(co-tunneling) coupling to a nearby Fermi-sea become prominent in
determining the spin dynamics in QDs.

Here we study the dynamics of an electron spin confined in a
self-assembled InAs/GaAs QD which is in turn embedded in a
Schottky heterostructure. In order to assess the relative
importance and external field dependence of the three elementary
spin-relaxation mechanisms we use the degree of resonant absorption as a measure: since the
degree of absorption i.e. strength of the trion-resonant light scattering is determined by competing optical spin pumping (OSP) \cite{AtatureS06} and spin relaxation induced by spin-reservoir interactions, the strength of spin relaxation can be inferred from the absorption measurement. First we demonstrate that at low magnetic fields (up to 1 Tesla) spontaneous spin-flip Raman scattering that allows for one-way
pumping into the optically-uncoupled spin-state is predominantly
mediated by a mixing between the electronic spin states induced by
the fluctuating hyperfine nuclear (Overhauser) field. Next, we show that upon
varying the external gate voltage by about 50 mV the spin relaxation due to exchange coupling to the
nearby Fermi-sea of electrons can be changed by as much as five
orders of magnitude. Finally, we show that at the high magnetic field regime (1 to 10 Tesla) spin pumping is due to heavy-light hole mixing and spin relaxation is dominated by phonons in conjunction with spin-orbit interaction.

Before proceeding we note that major advances in understanding
relaxation and decoherence of single confined electron spins have already been achieved in electrically defined QDs. By implementing a single-shot electrical read-out of a
QD spin Elzerman \textit{et al} have shown that spin lifetimes in
electrically defined QDs can reach up to $\sim$1 msec even at elevated magnetic fields of 8 Tesla \cite{ElzermanN04} and very recent measurements have revealed a relaxation time of 170ms at 1.75 Tesla \cite{Amasha06}. Similarly in double QDs Johnson \textit{et al} have investigated hyperfine-induced triplet-singlet relaxation \cite{JohnsonN05} Petta \textit{et al} have demonstrated coherent manipulation of singlet-triplet states \cite{PettaS05} and Koppens \textit{et al} have shown detection and control of hyperfine-induced singlet-triplet mixing \cite{KoppensS05} and Rabi oscillations using microwave pulses \cite{KoppensN06}. Further measurements on InAs/GaAs self-assembled QD ensembles have revealed $T_1$ times exceeding 20 msec at a magnetic field of 4 Tesla and a temperature of 1 Kelvin \cite{KroutvarN04}.

This paper is organized as follows: In section \ref{sec:spininteract} we 
introduce the coupling of the localized spin to nuclear spins, charge reservoir and phonons. Section \ref{sec:4leveltrion} then theoretically describes the  QD spin dynamics in the framework of the trion four-level system with spin-reservoir coupling. In section \ref{sec:spectrosc} we present our experimental results obtained with single QD absorption spectroscopy in distinct regimes of external electric and magnetic fields where different interactions dominate. Finally section \ref{sec:fullmap} gives an overview on the above-mentioned interactions together in a self-contained picture before the conclusions in section \ref{sec:conclusions}. For information about our sample structure and experimental
techniques we direct the reader to Appendix \ref{app:sample}.


\section{\label{sec:spininteract}Interactions of a Single Confined Spin}

\subsection{\label{ssec:nuclspinbasics}Nuclear Spins}

The interaction of a localized electron spin with a surrounding nuclear spin ensemble can be written in the form of the Fermi contact interaction which yields \cite{MerkulovPRB02}

\begin{eqnarray}
\label{eq:fermicontact}
\hat H_\textrm{hyp}=\frac{\nu_0}{8} \sum_{i} A_i |\psi(\bm R_i)|^2 ( \hat{\bm I_i} \cdot \hat{\bm \sigma} )
\end{eqnarray}
The sum runs over all nuclei $i$ in the lattice. $\nu_0$ is the volume of an InAs unit cell $\psi(\bm{R}_i)$ the electron envelope wavefunction at the $i$th nucleus and $\hat {\bm I_i}$ and $ \hat{\bm \sigma}$ are the spin operators of nuclear and electron spin. $A_i = (2 \mu_0 g_0 \mu_B \mu_i / 3 I_i) \left| u_c(\bm {R}_i) \right|^2$ is the hyperfine coupling strength and reflects the electron density described by the electron Bloch wavefunction $u_c(\bm {R}_i)$ at the site of the nuclei. $\mu_B$ is the Bohr magneton and $\mu_i$ the nuclear magnetic moment $\mu_0$ is the permeability of vacuum and $g_0$ the free-electron g-factor. 

In order to estimate the total number of nuclei within the spread of the electron wavefunction we use the dimensions of the QD: InAs and GaAs have Zincblende-type lattice with a lattice constant of 6.06 $\textrm \AA$ and 5.65 $\textrm \AA$. There are four Arsenic (nuclear spin I$_\textrm{As}$ = 3/2) and four Indium (I$_\textrm{In}$=9/2) or Gallium (I$_\textrm{Ga}$=3/2) atoms in a fcc unit cube. Taking this into account and assuming that the QD creates a box-like confinement that equals the dimensions of the QD ($\approx$ 20nm$\times$20nm$\times$5nm) the number of As atoms is on the order of 4$\times$10$^4$ and one has to add the same number of In or Ga atoms depending on the composition of the QD such that the total number of nuclei interacting with the QD spin can be taken to be $N=10^4$ to $10^5$.

We note that the Fermi contact interaction relies on a finite value of the Bloch wavefunction at the sites of the nuclei. Due to the s-like symmetry of their wavefunction electrons are susceptible to this interaction whereas the p-symmetric holes are not. For holes it is only possible to interact with nuclear spins via the much less efficient dipole-dipole interaction.

The hyperfine interaction constitutes a special case among the three spin-reservoir interactions discussed in this work as our experiments suggest that its dominant effect is not described by a \textit{textbook} system-reservoir interaction in the Born-Markov approximation as it is the case for phonon and exchange coupling. In the optical measurements presented here nuclear spins mainly act by exerting a quasi-static magnetic field (\textit{Overhauser field}) with rms-value on the order of $B_\textrm{nuc}$=15 mTesla. This field leads to a Rabi-type slowly varying coherent mixing of the spin ground states which can also be understood as precession of the electron spin in the nuclear magnetic field; in this context \textit{slowly varying} means that the correlation time of the hyperfine field  fluctuations ($\sim$1ms) is much longer than the precession time ($\sim$1ns) of the electron spin in this field. 

In contrast to the ground states the excited states remain unchanged as the hole is not susceptible to the nuclear magnetic field and the two electrons form a singlet which is immune to magnetic field variations too. The mixing of the ground states if strong enough would thus lead to a fluctuating observable splitting of the excitonic transitions however it turns out that its magnitude is less than the broadening of these transitions due to their radiative lifetime and therefore cannot be resolved in laser scanning absorption measurements even with experimental resolution much better than the transition linewidth. 

In order to understand the effect of the hyperfine field in optical experiments we consider two regimes: First an external magnetic field with strength smaller than the hyperfine field is applied or the external field is completely absent. Hence the direction of the total magnetic field seen by the electron spin is fully random after a nuclear field correlation time. As we will see in section \ref{sec:4leveltrion} fast bidirectional OSP will be the consequence inducing efficient spin relaxation dominating over other mechanisms as the one predicted in Ref.\cite{ErlingssonPRB02}. In the second regime the applied external field is much stronger than the hyperfine field. In this case the electron spin mainly sees the external magnetic field along the z-axis and the hyperfine field only leads to small fluctuations of the nuclear field vector. In this regime the light-induced spin relaxation is slow and other mechanisms such as phonons are dominant.

We will now analyze the effective hyperfine field in detail.
In the absence of dynamical nuclear spin polarization (DNSP) scenarios the action of nuclear spins upon the localized spin dominates the reverse action due to a much larger number of degrees of freedom on the side of the nuclear spins ensemble. Therefore the Hamiltonian (\ref{eq:fermicontact}) is reduced to an effective magnetic field seen by the QD spin which is commonly referred to as \textit{Overhauser field} 

\begin{eqnarray}
\label{eq:overhdef}
 \bm{B_N} = \frac{\nu_0}{8} \frac{\overline{A}}{g_e \mu_B} \langle \sum_{i} \hat{\bm{I}_i} \rangle
\end{eqnarray}
where $\overline{ A}$ is an average spin-nuclei coupling constant and $g_e$ is the QD electron g-factor.

For simplicity we will in the following treat the hyperfine field as a purely classical field $\bm{B_N}(t)$ with correlation time $\tau_{corr}\sim$1ms. The correlation time is expected to be similar to the decay time of nuclear spin polarization in the presence of a QD electron and absence of external magnetic field, as measured in \cite{Maletinskya07}. $B_\textrm{nuc}$ refers to the rms-value of the Gaussian distribution as defined by 

\begin{eqnarray}
f(\bm{B_N}) = \frac{1}{B_\textrm{nuc}^3 (2 \pi)^{3/2}} \exp(-|\bm{B_N}|^2/2 B_\textrm{nuc}^2)
\end{eqnarray}
which yields

\begin{eqnarray}
\label{eq:defnucgauss}
\langle  \bm{B_N}(t) \rangle &=& 0 \\ \nonumber
\langle |\bm{B_N}(t)|^2 \rangle  &=&  3 B_\textrm{nuc}^2
\end{eqnarray}
Here $\langle ~ \rangle$ denotes the time average over many correlation times. 

As our $\bm{B_N}(t)$ is quasi-classical, we treat the $B_{N,i}(t)$ with $i=x,y,z$ as independent random variables each one following a Gaussian distribution function given by

\begin{eqnarray}
\label{eq:bni_gauss}
f(B_{N,i}) = \frac{1}{B_\textrm{nuc} \sqrt{2 \pi}} \exp{\left( - \frac{B^2_{N,i}}{2 B_\textrm{nuc}^2}\right)}
\end{eqnarray}
and $f(\bm{B_N})$ is then given by the product of the Gaussian distributions of its three spatial components

\begin{eqnarray}
f(\bm{B_N}) &=&  \prod_{i=x,y,z} f(B_{Ni}) \end{eqnarray}
Clearly
\begin{eqnarray}
\langle B_{N,i}(t) \rangle &=& 0 \\ \nonumber
\langle |B_{N,i}(t)|^2 \rangle &=& B_\textrm{nuc}^2
\end{eqnarray}
$B_\textrm{nuc}$ can be written in the form (similar to \cite{MerkulovPRB02,Taylor2006})
\begin{eqnarray}
\label{eq:sigmabn}
B_\textrm{nuc} = \frac{b_0}{\sqrt{N}}
\end{eqnarray}
with $b_0$ a parameter characterized by the species of nuclei and the composition of the QD \footnote{$b_0^2/N=\frac{1}{3}(\frac{\nu_0}{8 g_e \mu_B})^2 \sum_i (A_i)^2 |\psi(R_i)|^4 \langle \bm {I}_i^2 \rangle = \frac{1}{3N} (\overline{A^2} ~ \overline{I(I+1)}) $ $/ (g_e \mu_B)^2$; this result is obtained by assuming a box-like confinement potential, further transforming the sum into an integral and using the localization volume of the QD electron given by $V_L^{-1}= \int \textrm d \bm r  |\psi(R_i)|^4 = 8/(N \nu_0).$ } and $N$ the number of nuclear spins interacting with the QD spin.

The QD composition is taken to be $90\%$ InAs and $10\%$ GaAs yielding $\overline{I (I+1)} = 13.2$ when averaging over the different nuclear species \cite{MaletinskyPRB07}. Similarly we obtain $\overline{A^2}=2500 \mu$eV$^2$ which yields $b_0= 3.0$ Tesla. Using (\ref{eq:sigmabn}) with N=$10^4$ to $10^5$ nuclear spins we obtain for our QDs
\begin{eqnarray}
\label{eq:bnexperimental}
B_\textrm{nuc} = \textrm{9.5 mT to 30 mT}
\end{eqnarray}
We can now rewrite (\ref{eq:fermicontact}) as

\begin{eqnarray}
\label{eq:h_overh}
\hat H_\textrm{Overh} &=& g_e \mu_B \bm{B_N}(t) \cdot \hat{\bm \sigma} 
\end{eqnarray}
Here the component of the nuclear field along the z-axis $B_{N,z}(t)$ only leads to Zeeman splitting whereas the in-plane components induce a mixing of the $|\uparrow\rangle$ and $|\downarrow\rangle$ states. The in-plane hyperfine field is
\begin{eqnarray}
\label{eq:bnxy}
B_{N,xy}^2(t) = B_{N,x}^2(t) + B_{N,y}^2(t) 
\end{eqnarray}
and we define 
\begin{eqnarray}
\label{eq:defomegah}
\hbar \Omega_H(t) = \frac{g_e \mu_B B_{N,xy}(t)}{2}
\end{eqnarray}
We note here that our measurement time (typically 10 to 100 ms) is longer than the correlation time of the nuclear field; i.e. for each measured data point we expect that we average over many configurations of the nuclear magnetic field \footnote{In this context we note that other experiments suggest that the dynamics of the nuclear spin ensemble could be altered by the measurement itself and the hyperfine field stays locked for a time on the order of seconds \cite{KoppensS05} Fig.4(b). In our measurements there is some evidence for alteration of dynamics at large magnetic fields and large gate voltage detunings (not shown in this work); however at low magnetic fields these locking effects do not seem to play a dominant role.}.


\subsection{\label{sec:cotunnel}Coupling to electron spin reservoir}

\begin{figure*}[ht]
	\includegraphics[angle = 0, width = \textwidth ]{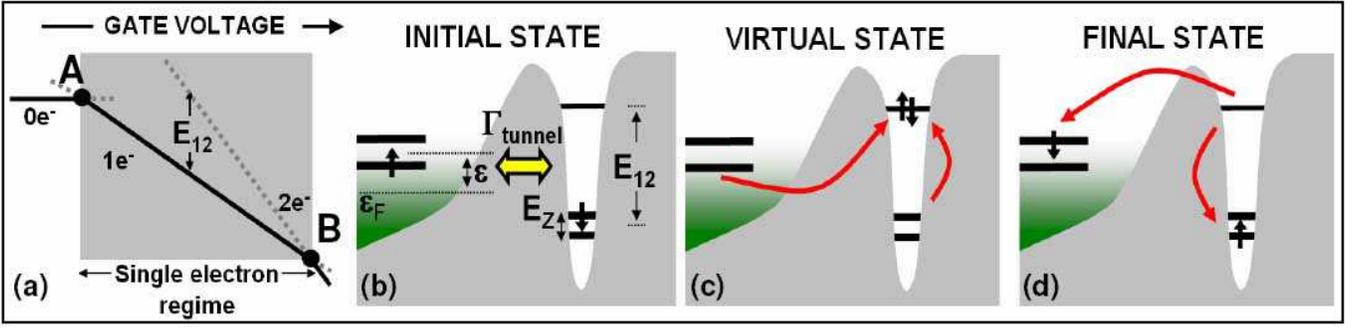}
	
	\caption{(Color online) (a) \textit{Stability diagram} of the QD ground states neglecting spin: Energies of the zero one and two-electron QD as a function of gate voltage. Crossover points are marked A and B. $E_{12}$ denotes the gate voltage-dependent energy difference between the singly-charged and the doubly-charged state or charging energy. (b to d) The QD can exchange its single electron with the charge reservoir via a virtual empty or two-electron (shown here) state. When one of the two singlet-electrons tunnels out it leaves the remaining QD spin in a mixed state equivalent to spin relaxation. $\Gamma_\textrm{tunnel}$ marks the tunneling rate through the 35-nm GaAs barrier $\varepsilon$ the detuning from the Fermi energy $\varepsilon_F$. $E_Z$ is the Zeeman splitting and $E_{12}$ the energy required to charge a second electron.}
	\label{fig:ctscheme}
\end{figure*}

Another interaction mechanism arises from the presence of the Fermi sea in the back contact (ref. to Appendix \ref{app:sample}) that couples to the QD via the tunneling barrier. It is well known that exchange interaction of a confined spin with an electron spin reservoir leads to co-tunneling \cite{AverinPRL90}\footnote{The type of co-tunneling relevant for us is inelastic co-tunneling as it leads to a transition between the spin ground states.} and at temperatures lower than the Kondo temperature $T_K$ to the formation of a Kondo singlet \cite{Goldhaber-GordonN98,CronenwettS98,GovorovPRB03,HelmesPRB05}. 

This interaction can be written as

\begin{eqnarray}
\label{eq:hcharge}
	\hat H_\textrm{charge}= \sum_{k,k'} \hbar g_{t,k} (e^\dagger_{k,\downarrow}  e^\dagger_{QD,\uparrow} e_{QD,\downarrow} e_{k',\uparrow} + c.c.)
\end{eqnarray}
where $e^\dagger_{QD,\sigma}$ and $e_{QD,\sigma}$ are the creation and destruction operators for an electron with spin $\sigma$ in the QD and similarly in the reservoir. $g_{t,k}$ is the tunneling matrix element, which is linked to the tunneling rate $\Gamma_\textrm{tunnel}$ by Fermi's Golden rule $\Gamma_\textrm{tunnel}=\frac{2 \pi}{\hbar} |g_{t,k}|^2 \rho(E)$ with $\rho(E)$ being the density of states in the back contact.

Fig. \ref{fig:ctscheme}(a) shows the energies of the empty, singly and doubly charged QD state as a function of gate voltage \cite{SeidlPRB05}. Which state has lowest energy obviously depends on the gate voltage and the QD attempts to reach it by either attracting or repelling electrons from or into the reservoir. Clearly there is a range of voltages (\textit{single electron charging plateau}) where it is energetically favorable for the QD to accomodate a single electron, marked by the shaded region in the figure. At the points A and B two charging levels are degenerate and fast exchange of the QD electron with the reservoir can take place, only limited by the tunneling rate. The \textit{real} gate voltages $V_A$ and $V_B$ that need to be applied in order to reach points A and B can vary from dot to dot depending on its confinement properties. 

In other words the QD is singly-charged for $V_A < V_g < V_B$ with $V_g$ the gate voltage. We define the plateau center $V_c$ 

\begin{eqnarray}
V_c = \frac{V_B - V_A }{2}
\end{eqnarray}
The gate voltage detuning is 
\begin{eqnarray}
\Delta V_g = V_g - V_c
\end{eqnarray}
The schematic co-tunneling process is depicted in Fig. \ref{fig:ctscheme}(b-d). The initial state is characterized by a QD with a single spin-down electron and Coulomb blockade prohibits tunneling of further electrons into the dot (b). Together with a spin-up electron from the reservoir a virtual spin singlet (c) is formed at energy difference $\Delta E =\varepsilon+E_{12}$ where $\varepsilon$ is the detuning from the reservoirs' Fermi energy $\varepsilon_F$. Finally the QD returns to the singly-charged state with a spin-up electron (d). 

$E_{12}$ and $E_{01}$ are given by

\begin{eqnarray}
E_{12} = E_2 - E_1 = e \frac{(V_B-V_g)}{\lambda} \\ \nonumber
E_{01} = E_1 - E_0 = e \frac{(V_g-V_A)}{\lambda}
\end{eqnarray}
with $E_i$ the energy of the QD charged with $i$ electrons and $\lambda$ a constant describing the geometric lever arm of the heterostructure.

Using (\ref{eq:hcharge}) one obtains for the cotunneling rate in second-order \cite{AverinPRL90,SmithPRL05}

\begin{eqnarray}
\label{eq:cotunnel}
  \kappa_\textrm{cotunnel} &=& \hbar \Gamma_\textrm{tunnel}^2 \int_\varepsilon \biggl| \frac{1}{\frac{e(V_g-V_A)}{\lambda}+ \varepsilon + \frac{\textrm i}{2}\hbar \Gamma_\textrm{tunnel} } + \\ \nonumber
 && \frac{1}{\frac{e(V_B-V_g)}{\lambda} - \varepsilon + \frac{\textrm i}{2} \hbar \Gamma_\textrm{tunnel} } \biggr|^2 f(\varepsilon)[1-f(\varepsilon)]\textrm d \varepsilon 
\end{eqnarray}
The integral is the sum over all second-order transitions with different detunings $\varepsilon$ from the Fermi energy according to Fig. \ref{fig:ctscheme}(b-d). In addition the term with $e(V_g-V_A)/\lambda=E_{01}$ describes the related process where the virtual state is an empty QD. $f(\varepsilon)$ is the Fermi function $f(\varepsilon)=1/(1+\exp(\varepsilon/kT))$. Expression (\ref{eq:cotunnel}) is valid under the condition $E_{Z,e} \ll kT$ i.e. for low magnetic fields. To obtain the exact expression for all magnetic fields the Fermi function terms in the integral have to be modified \footnote{\label{ft:fermif}With Zeeman splitting $\kappa_{\uparrow \rightarrow \downarrow}  \neq \kappa_{\downarrow \rightarrow \uparrow}$ i.e. the rate flipping the spin down is different from the rate flipping it up. The term $f(\varepsilon)[1-f(\varepsilon)]$ has to be replaced by $f(\varepsilon \mp \hbar \omega_z /2)[1-f(\varepsilon \pm \hbar \omega_z /2)]$ for $\kappa_{\downarrow \rightarrow \uparrow}$ and $\kappa_{\uparrow \rightarrow \downarrow}$ respectively.}.

The imaginary part of the denominator introduces a finite lifetime to the electronic states limited by the tunneling rate $\Gamma_\textrm{tunnel}$, implying that the main cause for broadening of the spin ground states is tunneling. This is relevant for elements of the integral with vanishing real part.

\begin{figure}[h]
	\includegraphics[scale=0.43]{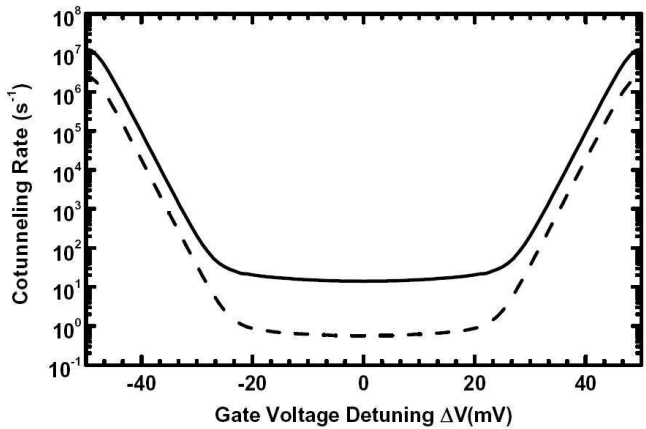}
	\caption{Expected cotunneling rate obtained using expression (\ref{eq:cotunnel}) with the parameters $\Gamma_\textrm{tunnel}$ = 0.1 ns$^{-1}$ (solid curve) and $\Gamma_\textrm{tunnel}$ = 0.02 ns$^{-1}$ (dashed curve), $V_A$ = -50 mV, $V_B$ = +50mV, kT = 300 $\mu$eV, $\lambda$=5.3. }
		\label{fig:cotcalc}
\end{figure}

In order to obtain an estimate for the cotunneling times in our structure 
we use results obtained on samples with 25nm tunneling barrier where in certain gate voltage regimes tunneling rate is larger than radiative recombination rate i.e. $\Gamma_\textrm{tunnel} > \Gamma$, leading to broadening in the linewidths observed in photoluminescence measurements \cite{SmithPRL05}. Then from a Wentzel-Kramers-Brillouin (WKB) estimation of the two different tunneling barriers together with the measured tunneling rate we estimate the tunneling rate $\Gamma_\textrm{tunnel}$ to be on the order of 0.02 to 0.1 ns$^{-1}$ in our structure. We take it to be independent of the gate voltage within the single electron regime.	Fig. \ref{fig:cotcalc} shows the calculated cotunneling rate obtained with expression (\ref{eq:cotunnel}) using two different tunneling rates of $\Gamma_\textrm{tunnel}$=0.02 ns$^{-1}$ and $\Gamma_\textrm{tunnel}$=0.1 ns$^{-1}$ representing the minimum and the maximum cotunneling rate we expect in our experiments respectively. Co-tunneling rate is characterized by its very non-linear voltage dependence. When close to the crossover points $V_A$ and $V_B$ it exhibits an ultra-steep slope; in contrast, the voltage-dependence is weak in the plateau center $V_c$.

\subsection{\label{sec:phonondecay}Spin - Phonon interaction}

It is known that spin relaxation in higher-dimensional systems is mainly due to spin-orbit (SO) interaction in conjunction with phonons \cite{D'yakonovSP-J86,KhaetskiiPRB00}. Despite being strongly suppressed, SO interaction is still an enabling mechanism for phonon-assisted spin flips in QDs and a considerable amount of theoretical work has been done on this spin relaxation mechanism
\cite{KhaetskiiPRB00,KhaetskiiPRB01,WoodsPRB02,GolovachPRL04,BulaevPRB05}. SO coupling is a well-known phenomenon in atomic physics as well as in semiconductors and is in general characterized by an interaction term of type $H_{SO} = \sum_{i,j} a_{ij} \hat{ l_i} \hat {\sigma_j}$ with $\hat{ \bm l}$ the angular momentum operator and $\hat {\bm \sigma}$ the spin operator of the electron; the sum runs over all pairs $i,j= x,y,z$. In the case of a crystal with bulk inversion asymmetry (BIA) such as GaAs and InAs SO coupling is of Dresselhaus type \cite{DresselhausPR55}. Similarly Rashba SO coupling results from asymmetry along the z-direction (SIA, structural inversion asymmetry) \cite{BychkovJoPCSSP84}. For a 2DEG the spin-orbit coupling can then be written as

\begin{eqnarray}
\label{eq:sodef}
H_{SO} = \beta (- p_x \sigma_x + p_y \sigma_y) + \alpha ( p_x \sigma_y - p_y \sigma_x)
\end{eqnarray}
where $\beta$ reflects the strength of the Dresselhaus SO coupling and $\alpha$ the Rashba SO interaction. In \cite{BulaevPRB05} the difference between the effect of Dresselhaus and Rashba SO coupling on QD spin relaxation are discussed. 

SO interaction leads to spin-orbital admixed states which can weakly couple to phonons leading to an effective spin-phonon reservoir coupling of type

\begin{eqnarray}
	\hat H_\textrm{ph,eff}=\hbar \sum_{q} g_\textrm{phon} (b^\dagger_{q}e^\dagger_{QD,\uparrow} e_{QD,\downarrow} + c.c.)
\end{eqnarray}
Here $\omega_z = c_s |q|$ with $c_s$ the speed of sound. The effective spin-phonon coupling $g_\textrm{phon}$ depends on the strength of SO interaction, the electron-phonon coupling strength as well as the phonon density of states at the Zeeman energy. 

The resulting spin relaxation rate is a function of magnetic field and is given by
\begin{eqnarray}
	\kappa_\textrm{phonon} = \frac{(g_e \mu_B B)^5}{\hbar (\hbar \omega_0)^4} \Lambda_p
\end{eqnarray}
where $\hbar \omega_0$ is the quantization energy for electrons and $\Lambda_p$ a dimensionless constant describing the strength of the piezoelectric coupling. The $B^5$ dependence valid for $E_{Z,e} \gg kT$ becomes replaced by $B^4 \cdot kT$ when $E_{Z,e} \ll kT$ due to the Boltzmann factor in \eqref{eq:relaxterms}, \cite{KhaetskiiPRB00,WoodsPRB02}.

In addition to this first mechanism there are ways of \textit{direct} spin-phonon coupling which turn out to be orders of magnitude weaker than the admixture mechanism described above \cite{KhaetskiiPRB00}. In \cite{KhaetskiiPRB01} the spin-flip rates due to different other phonon related mechanisms are estimated which all depend on electronic Zeeman splitting i.e. magnetic field as $\sim B^5$.

Beyond spin relaxation by emission or absorption of a single phonon, two-phonon processes in conjunction with SO interaction are predicted to dominate at small magnetic fields. In that case a phonon with wavevector $\bm p$ is scattered into a phonon with wavevector $\bm q$ with energy conservation $\hbar c_s |\bm{p-q}| = E_{Z,e}$. These two-phonon rates have characteristically strong temperature dependence, estimated to be in the range T$^7$ to T$^{11}$ \cite{KhaetskiiPRB01,WoodsPRB02}.

Alternatively it has been proposed that phonons together with the hyperfine-induced mixing of the Zeeman s-levels lead to relaxation of the QD spin. As already mentioned in the hyperfine subsection this mechanism is inefficient and the resulting rate is predicted to depend on the external magnetic field as $\sim B^3$ \cite{ErlingssonPRB02}: according to the calculations presented in this reference the rate will be less than $\kappa \sim$ 1s$^{-1}$ at a magnetic field of 1 Tesla when considering the larger quantization energy in our QDs.


\section{\label{sec:4leveltrion}Ground-state optical transitions of the singly-charged dot}

\subsection{Four-level model}

\begin{figure}[h]
	\includegraphics[scale=0.43]{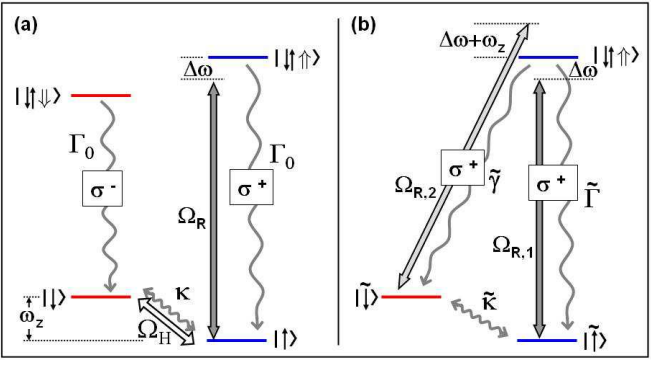}
	\caption{(a) Four-level system describing the singly-charged QD in magnetic field along the growth direction (z-axis). The electronic ground states with Zeeman splitting $\hbar \omega_z$  are vertically coupled by circularly polarized optical transitions to excitonic (trion) states. These consist of a heavy hole and two electrons forming a singlet. The fluctuations of the hyperfine field lead to a slowly varying coherent coupling $\Omega_H = \Omega_H(t)$ of the spin ground states. Incoherent spin-flip processes due to cotunneling and phonon-SO coupling are taken into account by relaxation rate $\kappa$. A laser is introduced at Rabi frequency $\Omega_R$ and detuning $\Delta \omega$ from the trion transition. (b) Transformed system after elimination of coherent coupling $\Omega_H$; this system is physically equivalent to that shown in (a). A weak hyperfine-induced diagonal transition appears at rate $\widetilde \gamma \propto \Omega_H^2 / B^2$. The laser is now detuned on the weak $\widetilde \gamma$ transition with a reduced Rabi frequency $\Omega_{R,2}$.}
		\label{fig:4level}
\end{figure}

A singly-charged QD is described as a four-level system with two ground states and two excited states, coupled by two vertical optical transitions, as shown in Fig.\ref{fig:4level}(a). The ground state $|\uparrow \rangle$ ($|\downarrow \rangle$) with angular momentum projection $m_z=+ 1/2$ ($m_z=-1/2$) is coupled to an excited state (trion state) formed out of two electrons in a singlet and a heavy hole $|\uparrow \downarrow  \Uparrow \rangle$ ($|\uparrow \downarrow  \Downarrow \rangle$) with spin projection $m_z = +3/2$ ($m_z = -3/2$), according to optical selection rules by $\sigma^+$ ($\sigma^-$) polarized optical transitions. 

The states are defined as

\begin{eqnarray}
\label{eq:defstates}
|\uparrow \rangle &=& e^{\dagger}_{QD, +1/2}  |0 \rangle \\ \nonumber
|\downarrow \rangle &=& e^{\dagger}_{QD, -1/2}  |0 \rangle \\ \nonumber
|\Uparrow \rangle &=& h^{\dagger}_{QD, +3/2}  |0 \rangle \\ \nonumber
|\uparrow \downarrow  \Uparrow \rangle &=& e^{\dagger}_{QD, -1/2} e^{\dagger}_{QD, +1/2} h^{\dagger}_{QD, +3/2}  |0 \rangle
\end{eqnarray}
where $e^{\dagger}_{QD, \sigma}$ ($h^{\dagger}_{QD, \sigma}$) is the operator that creates an electron (hole) in the QD with spin $\sigma$ along the z-axis and $|0 \rangle$ is the vacuum (empty dot) state.

All four states undergo different Zeeman shifts when an external DC magnetic field along the z-axis is applied, leading to Zeeman splitting of the optical transitions. A $\sigma^+$ polarized laser field is introduced at Rabi frequency $\Omega_R$ and detuning $\Delta \omega = \omega_0 - \omega_L$ with $\omega_0$ the frequency of the trion transition and $\omega_L$ the laser frequency. If only a $\sigma^+$ polarized laser field is present, the trion state with $m_z = -3/2$, i.e. ($|\uparrow \downarrow  \Downarrow \rangle$) is inactive since the coupling strength is reduced by a factor exceeding $10^3$ at magnetic fields larger than $60$mTesla, due to a combination of selection rules and, in the presence of a magnetic field, optical detuning \footnote{From the measured light polarization in our experiment we obtain, that the unwanted $\sigma^-$-component is suppressed by a $25\times$ factor. An additional factor comes from the optical detuning with respect to the diagonal transition in magnetic field, with the laser being on the strong transition. At $B=$60mTesla, the detuning leads to a further suppression $\sim 1/50$, altogether  suppressing the unwanted excitation of the weak transition by more than $1200\times$, increasing with magnetic field. At fields less or on the order of the hyperfine field, spin ground states are strongly mixed leading to all possible (vertical as well as diagonal) couplings in the four-level scheme, thus creating two differently polarized $\Lambda$-systems with equal decay rates to both ground states. The presence of the (unwanted and mainly suppressed) $\sigma^-$ polarized light would just increase the bidirectional OSP by a small amount.}. As we shall discuss shortly, the weak spontaneous emission to the other spin ground state cannot be neglected due to its long lifetime.

Thus, the system reduces to three levels; its quantum dynamics is fully described by the corresponding optical Bloch equations. These are obtained from a density matrix approach. 

The system Hamiltonian reads
\begin{eqnarray}
	\hat H= H_\textrm{Zeeman} + H_\textrm{int,rad} + \hat H_{\textrm{spin-reservoir}}
\end{eqnarray}
Then by tracing over the reservoir we obtain the master equation for the system (reduced) density operator $\hat \rho$

\begin{eqnarray}
\label{eq:mastereq}
\frac{\textrm d}{\textrm d t} \hat \rho = \frac{1}{\textrm i \hbar} [ \hat H_0, \hat \rho ] + \hat L_{\textrm{relaxation}}
\end{eqnarray}
The term $\hat H_0=H_\textrm{Zeeman} + H_\textrm{int,rad}$ describes the unitary dynamics and $\hat L_{\textrm{relaxation}}$ results from the interactions with reservoirs.

In the following we will discuss the different ingredients of this master equation. For the explicit optical Bloch equations we refer the reader to Appendix \ref{ssec:obe}. With a magnetic field along the z-axis $\bm{B_\textrm{ext}}= \bm{B_z} = (0,0,B_z)$, the total magnetic field at the QD is 

\begin{eqnarray}
\bm B = \bm{B_z} + \bm{B_N}
\end{eqnarray}
where the nuclear magnetic field (second term) is only seen by the electron spin, but not the hole spin. The Zeeman Hamiltonian then reads
using (\ref{eq:h_overh})

\begin{eqnarray}
 H_\textrm{Zeeman} &=&  H_{Z,e} +  H_{Z,h} \\ \nonumber
 H_{Z,e} &=& g_e \mu_B \hat{ \bm{B}} \cdot \hat{ \bm {\sigma}} \\ \nonumber
 &=&  \hbar   \Omega_H(t)  \hat \sigma_x + \hbar   \omega_z \hat\sigma_z  \\ \nonumber
 H_{Z,h} &=& g_h \mu_B B_z \cdot \sigma_z 
\end{eqnarray}
with $\Omega_H(t)$ as defined in (\ref{eq:defomegah}). In addition,

\begin{eqnarray}
\hbar \omega_z = g_e \mu_B (B_z+ B_{N,z}(t))
\end{eqnarray}
The interaction with the radiation field in semi-classical form is

\begin{eqnarray}
	H_\textrm{int,rad} = \hbar \Omega_R \left(  e^{\textrm i \Delta \omega t}  e^{\dagger}_{QD, -1/2} h^{\dagger}_{QD, +3/2}  + \textrm{h.c.} \right)
\end{eqnarray}
Here, the laser detuning is $\Delta\omega=\omega_0-\omega_L$ and the Rabi frequency is $\Omega_R$. For simplicity we use the notation

\begin{eqnarray}
\label{eq:oldbasis}
| 1 \rangle &=& | \downarrow \rangle\\ \nonumber
| 2 \rangle &=& | \uparrow \rangle\\ \nonumber
| 3 \rangle &=& | \uparrow \downarrow \Uparrow \rangle
\end{eqnarray}
$H_0$ can be written as

\begin{eqnarray}
\hat H_0 = \hbar \begin{pmatrix}
\omega_z & \Omega_H(t) & 0 \cr
\Omega_H(t) & 0 & \Omega_R \cr 
0 & \Omega_R & \omega_0-\omega_L \cr
\end{pmatrix}
\end{eqnarray}
We note here that a separation of timescales, i.e. $\Omega_H(t)\approx \Omega_H$ can be done, since the time evolution of $\Omega_H(t)$ is much slower than all the timescales over which the system reaches steady-state.

$\hat H_{\textrm{spin-reservoir}}$ refers to the interaction of the QD spin with the thermal reservoirs of electron spins and phonons.

\begin{eqnarray}
\label{eq:hamilspinres}
	\hat H_{\textrm{spin-reservoir}}= \hat H_\textrm{charge} + \hat H_\textrm{ph,eff}
\end{eqnarray}
In our three-level model, each spin-reservoir coupling is treated as an incoherent relaxation rate $\kappa_i$ coupling states $|1\rangle \leftrightarrow |2\rangle$ bidirectionally. The total rate of spin relaxation, identical with the inverse of the spin $T_1$ time ($\kappa^{-1} = T_1$), is the sum of all contributions $\kappa_i$. We note that all $\kappa_i$ depend on external magnetic and/or electric field.

\begin{eqnarray}
\label{eq:kappasum}
	\kappa &=& \sum \kappa_i = f(V_g,B_\textrm{ext}) \\ \nonumber
 &=& \kappa_\textrm{cotunnel}(V_g,B_\textrm{ext}) + \kappa_\textrm{phonon}(B_\textrm{ext}) + \kappa_\textrm{exp}
\end{eqnarray}
Here, we have included a term that describes an experimentally induced spin relaxation rate $\kappa_\textrm{exp}$. Unless specified otherwise, this relaxation is absent, but can be invoked by large-amplitude gate voltage modulation in electron cycling experiments as discussed in section \ref{sec:spectrosc}.

After adding the relaxation terms due to the coupling to the thermal bath of radiation field modes (spontaneous emission terms) at rate $\Gamma$, the relaxation terms in the Lindblad form are \cite{Yamamoto99} 

\begin{eqnarray}
\label{eq:relaxterms}
\hat L_{\textrm{relaxation}}= &\frac{\Gamma}{2}& (2 \hat\sigma_{23} \hat\rho\hat \sigma_{32} - \hat\sigma_{33}\hat\rho - \hat\rho \hat\sigma_{33}) 
\\ \nonumber + &\frac{\kappa}{2}& \bar n (2 \hat\sigma_{12} \hat\rho\hat \sigma_{21}- \hat\sigma_{22}\hat\rho - \hat\rho\hat \sigma_{22}) 
\\ \nonumber + &\frac{\kappa}{2}& (\bar n +1) (2 \hat\sigma_{21} \hat\rho\hat \sigma_{12}-\hat \sigma_{11}\hat\rho -\hat \rho \hat\sigma_{11})
\end{eqnarray}
Here, $\hat \sigma_{ab} = |a \rangle \langle b|$ is the projection operator, $\Gamma$ is the spontaneous radiative decay rate of the optical transition and $\kappa$ the total spin relaxation rate. At temperatures smaller or comparable to the electronic Zeeman splitting $kT < E_{Z,e}$, a Boltzmann factor $\bar n=1/(\exp(g_e \mu_B B/kT)-1)$ needs to be taken into account which leads to thermalization of the electron spin, i.e. in the absence of light $\rho_{11} /  \rho_{22}  = \exp(-E_{Z,e}/kT)$, where $E_{Z,e}$ is the electronic Zeeman energy. In the case of exchange coupling, the $\bar n$ terms cannot be regarded as an \textit{occupancy}; it can however be shown that a similar factor appears in the co-tunneling rate \eqref{eq:cotunnel} when Zeeman splitting is taken into account \footnote{Including Zeeman splitting, relation (\ref{eq:cotunnel}) can be written in the form $\kappa_{\uparrow \rightarrow \downarrow} = \exp(-\hbar \omega_z /(2 kT) ) \int_\varepsilon \textrm d \varepsilon A(\varepsilon) \tilde f(\varepsilon)$ and similarly 
  $\kappa_{\downarrow \rightarrow \uparrow}= \exp(\hbar \omega_z /(2 kT) ) \int_\varepsilon \textrm d \varepsilon A(\varepsilon) \tilde f(\varepsilon)$ with $\tilde f(\varepsilon)= [ (1+\exp(\varepsilon+\hbar \omega_z /(2 kT))) (1+\exp(\varepsilon-\hbar \omega_z /(2 kT))) ]^{-1}$. In (\ref{eq:relaxterms}), the $\kappa \bar n$ term has to be replaced by $\kappa_{\uparrow \rightarrow \downarrow}$, whereas the $\kappa (\bar n+1)$ is replaced by $\kappa_{\downarrow \rightarrow \uparrow}$. As a consequence, the ratio of spin-up versus spin-down state occupation is governed, as one would expect, by the Boltzmann factor $\kappa_{\uparrow \rightarrow \downarrow}/\kappa_{\downarrow \rightarrow \uparrow}= \exp(-\hbar \omega_z /(kT)$}.

We note, that coupling of the trion states due to hole spin relaxation has been neglected here. This issue is discussed in section \ref{sec:phononexp}.

\subsection{Dressed states and rate equation description of spin pumping}

In order to gain a better understanding of our 3-level system, in the following we will transform the system into another basis, that gives an intuitive picture and we will see that the 3-level system eventually is a $\Lambda$-system. With some approximations, this allows us to capture the main features of the spin dynamics in the form of rate equations. 

The new basis is:
\begin{eqnarray}
\label{eq:newbasis}
| \widetilde \downarrow \rangle =  | \widetilde 1 \rangle &=& \cos \phi | 1 \rangle - \sin \phi | 2 \rangle \\ \nonumber
| \widetilde \uparrow \rangle = | \widetilde 2 \rangle &=& \sin \phi | 1 \rangle + \cos \phi | 2 \rangle \\ \nonumber
| \widetilde {\uparrow \downarrow \Uparrow } \rangle = | \widetilde 3 \rangle &=& | 3 \rangle
\end{eqnarray}
with $\phi = \Omega_H / \omega_z$. In order to simplify the calculations we put the additional constraint of $\phi \ll 1$ and only take into account first-order terms in $\phi$. For details we refer to Appendix \ref{ssec:sytrafo}.

The transformed Hamiltonian is

\begin{eqnarray}
\label{eq:newhamil}
\widetilde H_0 = \hbar \left(  \begin{array}{ccc}
\omega_z & 0  & \Omega_{R,1} \\
0 & 0  &  \Omega_{R,2}\\
\Omega_{R,1} & \Omega_{R,2}& \Delta \omega
\end{array} \right) 
\end{eqnarray}

The off-diagonal terms due to $\Omega_H$ have been eliminated and it is obvious from the Hamiltonian (\ref{eq:newhamil}) that both ground states couple to the excited state via an optical transition. Also the spontaneous emission terms become modified into a strong and a weak channel, marked by spontaneous emission rates $\widetilde \Gamma$ and $\widetilde \gamma$. For details refer to Appendix \ref{ssec:sytrafo}.

The result of the transformation is shown in Fig.\ref{fig:4level}(b): A single laser that interacted with the $\sigma^+$  trion transition is now represented by two laser fields coupling states $|\widetilde 1 \rangle$ and $|\widetilde 3 \rangle$ ($|\widetilde 2 \rangle$ and $|\widetilde 3 \rangle$, respectively). Effectively, the system can be decomposed into two two-level systems with its own spontaneous emission rates, Rabi frequencies and effective laser detunings. Those are for the $|\widetilde 1\rangle \leftrightarrow |\widetilde 3\rangle$ subsystem:

\begin{eqnarray}
\label{eq:newlaser11}
\widetilde \gamma &=& \phi^2 \Gamma \\ \nonumber
\widetilde\Omega_{R,1} &=& \phi \Omega_R \\ \nonumber
\widetilde\Delta \omega_1 &=& \Delta \omega + \omega_z  
\end{eqnarray}
And for the $|\widetilde 2\rangle \leftrightarrow |\widetilde 3\rangle$ subsystem:
\begin{eqnarray}
\label{eq:newlaser2}
\widetilde \Gamma &=& \Gamma \\ \nonumber
\widetilde\Omega_{R,2} &=& \Omega_R \\ \nonumber
\widetilde\Delta \omega_2 &=& \Delta \omega 
\end{eqnarray}
It is clear that, in this first order approximation, the couplings of the dressed $| \widetilde 2\rangle \leftrightarrow |\widetilde 3\rangle$ system are the same as for the bare $| 2\rangle \leftrightarrow |3\rangle$ transition. 

In the following we will discuss the properties of the transformed three-level system with a resonant laser on the $|\widetilde2\rangle \leftrightarrow |\widetilde3\rangle$ subsystem, i.e. under the condition $\Delta \omega=0$.

For that, we start out with the system being in state $|\widetilde 2 \rangle$ and we ignore state $|\widetilde 1 \rangle$. After an intermediate time $t_0$ given by $\widetilde \gamma^{-1} \gg t_0 \gg \Gamma^{-1}$, the laser field induces a steady-state occupation of the excited state $\widetilde \rho_{33}(t_0)$. Further, for times much longer than $(\widetilde \rho_{33}(t_0)\widetilde \gamma)^{-1}$ the system can also be found in state $|\widetilde 1 \rangle$.

The net effect of this spin-flip Raman process is a transfer of occupation from state $|\widetilde 2\rangle$ to state $|\widetilde 1\rangle$. We will refer to this process as OSP, due to its similarity to experiments performed with atoms \cite{BrosselJDPELR52}. Further, we note that a scheme that uses OSP for spin state preparation had been proposed in \cite{ShabaevPRB03}, and state preparation has been experimentally demonstrated in \cite{AtatureS06}.

In order to obtain the OSP rate that transfers the system from state $|\widetilde 2\rangle$ to state $|\widetilde 1\rangle$ under the presence of a resonant laser ($\Delta \omega =0$), we assume the system is in state $|\widetilde 2\rangle$. As already mentioned, $\widetilde \gamma$ represents a weak escape channel only, and we can work with the steady state occupation for the trion state $|\widetilde 3\rangle$, which is \cite{Loudon03}

\begin{eqnarray}
\label{eq:rho33ststate}
\widetilde \rho_{33}(t=\infty) = \frac{\Omega_{R,2}^2}{\widetilde \Gamma^2 + 2 \Omega_{R,2}^2} 
\end{eqnarray}
The OSP rate from state $|\widetilde 2 \rangle$ to $|\widetilde 1 \rangle$ then is the trion steady state occupation times the spontaneous emission rate into the weakly allowed channel:

\begin{eqnarray}
\label{eq:rho33}
R_{2 \rightarrow 1}=\widetilde  \rho_{33}(t=\infty) \cdot \widetilde \gamma
\end{eqnarray}
The $|\widetilde 1\rangle \leftrightarrow |\widetilde 3\rangle$ sub-system has a spontaneous emission rate of $\widetilde \gamma$, however the relaxation of its excited state is governed by a strong escape channel which is determined by the $\widetilde \Gamma$ rate. The rate of transfer from $|\widetilde 1\rangle$ to $|\widetilde 2\rangle$ is \cite{Loudon03}

\begin{eqnarray}
\label{eq:12osp1}
R_{1 \rightarrow 2} =   \frac{\Omega_{R,1}^2 \widetilde \Gamma} {4 \omega_z^2 + \widetilde \Gamma^2} \approx \frac{\Omega_{R,1}^2 \widetilde \Gamma} {4 \omega_z^2}
\end{eqnarray}
Using (\ref{eq:newlaser11}) to (\ref{eq:12osp1}), the time-averaged ratio of ground state occupations is obtained under the condition that $\kappa \ll R_{1 \rightarrow 2}, R_{2 \rightarrow 1}$

\begin{eqnarray}
\label{eq:pop1122}
\frac{\widetilde \rho_{22}}{\widetilde \rho_{11}} = \frac{R_{1 \rightarrow 2}}{R_{2 \rightarrow 1}}= \frac{\widetilde \Gamma^2 + 2 \Omega_R^2}{4 \omega_z^2 + \widetilde \Gamma^2}   \approx \frac{\widetilde \Gamma^2}{4 \omega_z^2} 
\end{eqnarray}
The approximation on the right hand side is valid in the limit of a weak incident beam ($\Omega_R \ll \widetilde \Gamma$) and a Zeeman splitting largely exceeding the trion decay rate ($\omega_z \gg \widetilde \Gamma$). Under these conditions, from (\ref{eq:rho33ststate}) follows that $\widetilde \rho_{33}(\infty) \ll 1$, such that together with $\textrm{tr} (\widetilde \rho) =1$ we can safely assume

\begin{eqnarray}
\label{eq:sumrho}
\widetilde \rho_{11} +\widetilde \rho_{22} \approx 1
\end{eqnarray}
and we obtain
\begin{eqnarray}
\label{eq:rho22steadystate}
\widetilde \rho_{22} (t = \infty) \approx \frac{1}{1+\frac{4 \omega_z^2}{\tilde \Gamma^2}}
\end{eqnarray}
In the case $R_{1 \rightarrow 2} \ll \kappa, R_{2 \rightarrow 1}$ the ratio of ground state occupations is
\begin{eqnarray}
\label{eq:pop1122largek}
\frac{\widetilde \rho_{22}}{\widetilde \rho_{11}} = \frac{\kappa}{R_{2 \rightarrow 1} + \kappa}
\end{eqnarray}
and together with (\ref{eq:sumrho}) we obtain 
\begin{eqnarray}
\label{eq:pop22largek}
\widetilde \rho_{22} (t = \infty) &\approx& \frac{1}{2+\zeta} \\ \nonumber
\zeta &=& \frac{\tilde \gamma}{\kappa} \cdot \frac{\Omega_R^2}{\tilde \Gamma^2 + 2 \Omega_R^2}
\end{eqnarray}
Hence in the case of fixed laser intensity i.e. constant $\Omega_R^2$, the spin-state occupations are determined by the ratio of OSP rate versus spin relaxation rate.

\subsection{Hole mixing}
\label{ssec:holemix}

Valence-band mixing, as described by the Luttinger Hamiltonian \cite{LuttingerPR56}, is a well-known feature in quantum wells. Similarly, it is expected to play a role in quantum dots. With valence-band mixing, a heavy hole acquires a small contribution of light holes and vice versa such that the effective hole state as it was defined in (\ref{eq:defstates}) has the form

\begin{eqnarray}
|\Uparrow \rangle_\textrm{hmix} &=& (h^{\dagger}_{QD, +3/2} \\ \nonumber
&& ~~ + \varepsilon_+ h^{\dagger}_{QD, +1/2} +  \varepsilon_- h^{\dagger}_{QD, -1/2}) |0 \rangle
\end{eqnarray}
with $|\varepsilon_\pm| \ll 1$. Pseudopotential calculations for self-assembled InAs QDs yield admixtures on the order of a few percent \cite{BesterPRB03}. As it has been pointed out in \cite{CalarcoPRA03} valence-band mixing would have a major impact upon the effective optical selection rules by introducing a diagonal relaxation channel between states $|3\rangle$ and $|1\rangle$ due to the admixed light hole component of state $|3\rangle$. 
Two cases have to be distinguished: First, the mixing contribution associated with $\varepsilon_+$ further leads to an effective coherent laser coupling in addition to the coupling induced by hyperfine interaction ($\propto \frac{\Omega_H}{\omega_z} \Omega_R$) at a detuning $\Delta \omega + \omega_z$ as shown in Fig. \ref{fig:4level}(b). Second, the $\varepsilon_-$ part essentially only appears as a relaxation channel without coherent laser coupling, as the dipole moment of this linearly polarized transition lies along the propagation axis of the laser beam and therefore cannot be excited. Hence the following diagonal relaxation terms are added to equation (\ref{eq:relaxterms}) \footnote{This can be seen from $ \langle \uparrow \downarrow  \Uparrow_\textrm{hmix} | H_\textrm{int,rad} | \downarrow \rangle  \neq 0 $ as opposed to $\langle \uparrow \downarrow  \Uparrow | H_\textrm{int,rad} | \downarrow \rangle  = 0 $} 

\begin{eqnarray}
\hat L_{\textrm{relax,hm}}= \frac{\gamma_\textrm{hm}}{2} (2 \hat\sigma_{13} \hat\rho\hat \sigma_{31} - \hat\sigma_{33}\hat\rho - \hat\rho \hat\sigma_{33}) 
\end{eqnarray}
with $\gamma_\textrm{hm} = |\varepsilon|^2 \Gamma = (|\varepsilon_+|^2+|\varepsilon_-|^2) \Gamma$.

This diagonal rate leads to OSP in a way similar to the $\widetilde \gamma$ channel enabled by hyperfine interaction. The main difference between hyperfine-induced OSP and valence-band mixing induced OSP is that the first one is magnetic field dependent as discussed previously, and the latter is not: the valence-band mixing strength $\varepsilon$ is expected to be independent of magnetic field, as long as the Zeeman splitting is much smaller than the heavy-light hole splitting ($\Delta_{hl} >$10meV), which is true for all realistic experimental magnetic fields. Since the hyperfine-induced OSP rate drops with magnetic field $\widetilde \gamma \propto B^{-2}$, hole-mixing induced OSP should dominate at high fields. From our measurements at high magnetic fields (Fig.\ref{fig:sophonon}(b)) we extract a $\gamma_{hm}^{-1}$ of $2\pm 0.8 \mu$s which yields a hole mixing strength of $|\varepsilon | \sim 2.2\%$. This $|\varepsilon |$ value is indeed much smaller than 1, but we also expect the exact value to vary from one QD to another. 

We note that a slightly tilted external magnetic field would yield identical dynamics in the absence of any hole mixing since it would lead to mixing of electronic states induced by the in-plane component of the applied field. These two fundamentally different mechanisms are experimentally indistinguishable for a fixed magnetic field orientation. Therefore we repeated our experiments as a function of sample tilt under a magnetic field. For a $\pm 1.5^\circ$ coverage of tilt in all directions our measurements yielded no observable change in the measured quantity $\gamma_{hm}$. Hence, we can safely state that the inherent hole mixing in our QDs indeed dominates over small-angle tilt-induced mixing of electronic spin states.

\section{Single dot absorption spectroscopy with resonant laser}
\label{sec:spectrosc}
\subsection{Experimental method}

All data shown in this work has been obtained using differential transmission (DT) technique \cite{KarraiSAM03,HogeleDEZ05,AlenAPL06,AlenAPL03,HogelePRL04,HogelePE04}. A narrow-band laser is scanned over the QD resonance while the transmitted light intensity is measured by photodetectors (for details refer to Appendix \ref{app:sample}). A QD transition then results in a transmission dip i.e. absorption due to resonant Rayleigh scattering. In \cite{AlenAPL03} the DT technique is used to investigate fine structure of the $X^0$ which exhibits a single two-level system in the case that the scanning laser is linearly polarized along one of the QD axes and only one transition of the $X^0$ doublet is addressed. 

In order to link the experimentally observable absorption i.e. intensity drop on the detectors to the 3-level system of the singly-charged QD, we use the effective $\Lambda$-system picture as described in the previous section. With a resonant laser and large external magnetic field along the z-axis i.e. $\Delta \omega=0, ~ \omega_z \gg \tilde \Gamma \gg \tilde \gamma$, the $2 \leftrightarrow 3$ subsystem with strong spontaneous emission rate $\tilde \Gamma$ acts as the main scattering source. In the $B \rightarrow \infty$ limit only the strong transition contributes to light scattering. 

The QD response in this type of DT experiments is discussed in the above mentioned references \cite{AlenAPL06,AlenAPL03}. In the following we therefore only sketch the link between our 3-level system and the intensity of the light transmitted through the sample.

In general, for a given two-level system with ground state $|g\rangle$, excited state $|e\rangle$ and spontaneous emission rate $\Gamma$ the linear susceptibility $\chi = \chi^{(1)}_\textrm{QD}$ as defined by $P=\varepsilon_0 \varepsilon \chi E$, with $P$ the polarization describing the QD dipole, $E$ the electric field, $\varepsilon_0$ the dielectric constant of vacuum and $\varepsilon$ the dielectric constant of medium (GaAs). $\chi$ can be written in terms of the steady-state off-diagonal density matrix element $\rho_{eg} (\infty)$

\begin{eqnarray}
\label{eq:suscept}
\chi(\omega) &=& \frac{2 \pi c^3 v}{\omega_0^3} \frac{ \frac{\Gamma}{2} (\omega_0 - \omega + \textrm i \Gamma/2)}{(\omega_0 - \omega)^2 + \frac{\Gamma^2}{4} + \Omega_R^2/2} \\ \nonumber
&=& - \frac{2 \pi c^3 v}{\omega_0^3}  \frac{\Gamma}{\Omega_R}~ \rho_{eg} (\infty)
\end{eqnarray}
with $c$ the speed of light in vacuum, $\omega_0$ the frequency of the optical transition, $\omega$ the laser frequency and $v$ the scattering volume of the QD. The real part of $\chi$ leads to a dispersive QD response to the laser detuning whereas the imaginary part produces an absorptive Lorentzian response as can be read from (\ref{eq:suscept}).

The signal detected in a DT experiment arises from interference of the forward scattered field together with the excitation field \cite{HogeleDEZ05}. This can also be seen from the optical theorem \cite{NewtonAJoP76} which relates the \textit{absorption} cross-section of the dipole to the forward-scattering amplitude. When the QD is exactly at the focus, the imaginary part of the scattered field is in phase with the excitation laser.

After collecting all factors describing spatial mismatch between excitation field and QD scattering cross-section we define relative absorption as

\begin{equation}
\label{eq:deftheta}
	\Theta(\Delta \omega) =1- \frac {T(\Delta \omega)}{T_\textrm{off}} = s \cdot  \textrm{Im} (- \frac{\Gamma}{\Omega_R} \rho_{eg}(\infty))
\end{equation}
where $T(\Delta \omega)$ refers to the transmitted intensity as a function of laser detuning and $T_\textrm{off}$ to the transmitted intensity far off resonance in the limit ($\Delta \omega \rightarrow \infty$). On resonance in the weak excitation regime i.e. $\Omega_R \ll \Gamma$ the term Im$(\frac{\Gamma}{\Omega_R} \rho_{eg}(\infty))$ reaches 1 and $s$ is a scaling factor characterizing the maximum theoretical absorption contrast that is given by

\begin{equation}
 s = S_\textrm{exp} \frac {\sigma_0}{A_\textrm{L}}
\end{equation}
valid for a weak focusing geometry. Here $\sigma_0 = (3/2\pi) (\lambda/n)^2$ is the scattering cross-section of the two-level system in the weak excitation limit and $A_\textrm{L}$ is the laser spot area. The factor $S_\textrm{exp}$ accounts for reduction of signal due to our lock-in detection scheme and experimental imperfections.
Further, with $\Delta \omega =0$

\begin{equation}
\rho_{eg}(\infty) = - \textrm i \frac{\Gamma}{\Omega_R} \rho_{ee}(\infty)
\end{equation}
and equation (\ref{eq:deftheta}) reads 

\begin{equation}
\label{eq:deftheta2}
	\Theta(\Delta \omega=0) =1- \frac {T(\Delta \omega=0)}{T_\textrm{off}} =   s \cdot \frac{\Gamma^2}{\Omega_R^2}  \rho_{ee}(\infty)
\end{equation} 
We can now apply this scheme to our double 2-level system and we will assume in the following that the laser is on resonance with the strong trion transition i.e. the $| \widetilde 2 \rangle \leftrightarrow | \widetilde 3 \rangle$ subsystem. Clearly, $\widetilde \rho_{33} \equiv \rho_{ee}$ and $\widetilde \rho_{23} \equiv \rho_{ge}$. When $\omega_z \gg \Omega_H$ the weak transition does not contribute to the absorption signal. Using (\ref{eq:deftheta2}) we can infer the value of the spin-up state occupation $\widetilde \rho_{22}(\infty)$ from our absorption measurements when varying parameters such as magnetic field and gate voltage but keeping laser power constant i.e. constant $\Omega_R$.

We still need a calibration point, i.e. an experimental value of $\Theta(0)$ for a known $\widetilde \rho_{22}(\infty)$. In the absence of an external magnetic field the spin ground states can be considered to be fully mixed due to the in-plane part of the Overhauser field leading to $\widetilde \gamma \sim \Gamma$ and a branching ratio of $\eta = 1$. As a consequence the $|\widetilde 1\rangle$-$|\widetilde 3\rangle$ and the $|\widetilde 2\rangle$-$|\widetilde 3\rangle$ transitions equally contribute to light scattering and fast bidirectional OSP takes place, leading to a fully randomized spin i.e.

\begin{equation}
\label{eq:thcalibration}
\widetilde \rho_{11}(t=\infty,B=0) = \widetilde \rho_{22}(t=\infty,B=0) =\frac{1}{2}
\end{equation}
Given the relations (\ref{eq:deftheta2}) and (\ref{eq:thcalibration}) and the fact that the absorption measurement constitutes a relative  measurement of the steady-state occupation $\widetilde \rho_{22}(\infty)$ and expression (\ref{eq:deftheta2}) can be rewritten by replacing the $s$-factor
\begin{equation}
\label{eq:deftheta3}
	\Theta(\Delta \omega=0) =1- \frac {T(\Delta \omega=0)}{T_\textrm{off}} =   s' \cdot \frac{\Gamma^2}{\Omega_R^2}  \widetilde \rho_{22}(\infty)
\end{equation} 
The $s'$-factor can be experimentally determined using (\ref{eq:thcalibration}).

\subsection{\label{ssec:spinpump}Optical spin pumping}

\begin{figure}[h]
	\includegraphics[width=0.45 \textwidth]{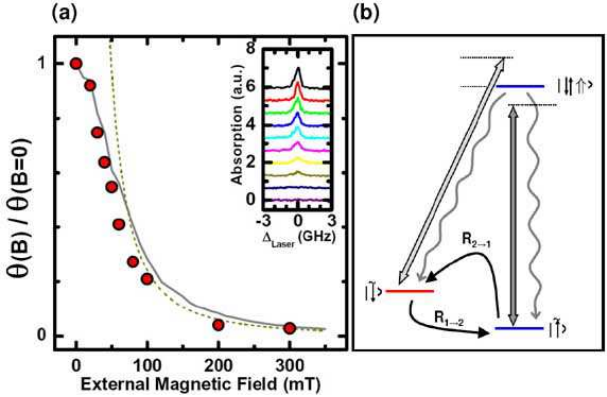}
	\caption{(Color online) (a) Absorption maxima in the plateau center plotted as a function of magnetic field $B_z$. A drop occurs with increasing $B_z$ due to OSP which, at low magnetic fields, dominates over co-tunneling and phonon interaction. The grey line is a numerical simulation using $\Omega$ = 0.6 $\Gamma$, $B_\textrm{nuc}$=15mT, $\Gamma^{-1}$=0.8ns, additional diagonal relaxation $ \gamma_{hm}^{-1}=2 \mu$s, $\kappa^{-1}$=10ms. The inset shows the corresponding raw laser scans from 0 T (top) to 300 mT (bottom). The peaks have been shifted laterally to eliminate Zeeman splitting. (b) Optically induced spin pumping rates $R_{2 \rightarrow 1}$, transferring the system into the dark state, and $R_{1 \rightarrow 2}$ the back-pumping rate.}
	\label{fig:absdrop}
\end{figure}

Fig.\ref{fig:absdrop}(a) shows  absorption on resonance on the blue Zeeman transition as a function of magnetic field normalized to on-resonance absorption at 0 Tesla i.e. $\Theta(B_\textrm{ext})/\Theta(B_\textrm{ext}=0)$. The gate voltage was kept in the plateau center i.e. in a regime where $\kappa_\textrm{cotunnel}$ is minimal. The inset shows the corresponding raw laser scans for 0 T (top) to 300 mT (bottom). The zero positions of the probe laser detuning has been readjusted in the graphs to compensate the Zeeman splitting. 

Absorption drops by nearly two orders of magnitude over the plotted range of $B_z$ = 0 to 300 mT. With a resonant laser in the weak excitation limit and Zeeman splitting much larger than the trion transition linewidth i.e. $\Delta \omega=0, ~ \omega_z \gg \tilde \Gamma \gg \tilde \gamma$ equation (\ref{eq:rho22steadystate}) yielded

\begin{eqnarray}
\nonumber \widetilde \rho_{22} (t = \infty) \approx \frac{1}{1+\frac{4 \omega_z^2}{\tilde  \Gamma^2}}
\end{eqnarray}
all provided that the spin relaxation rate $\kappa \ll R_{2 \rightarrow 1},R_{1 \rightarrow 2}$ which we can safely assume for low magnetic fields \cite{ErlingssonPRB02,KhaetskiiPRB01} and strongly suppressed exchange coupling in the gate voltage plateau center. Consistent with (\ref{eq:rho22steadystate}), drop of absorption follows a $B^{-2} \propto \omega_z^2$ law indicated by the dashed line. For fields less than 100mT (see Fig. \ref{fig:absdrop}(a)) the approximations included in (\ref{eq:rho22steadystate}) do not hold any more and $\tilde \Gamma \sim \tilde \gamma$. Without any approximation the steady-state solutions of the optical Bloch equations are evaluated (solid line) numerically; they are in excellent agreement with our data at all magnetic fields. The Rabi frequency $\Omega$ in units of $\Gamma$ for a given incident laser power can be independently determined by saturation spectroscopy and power broadening measurements. The radiative lifetime $\Gamma^{-1}$= 0.8ns used in our simulation is based on a measurement in as-grown dots \cite{DalgarnoAPL06}.

\subsection{\label{ssec:ecycling}Electron cycling}

Given that the OSP rates $R_{1 \rightarrow 2}$, $R_{1 \rightarrow 2}$ and the spin relaxation rate $\kappa$ are unknown the experimental data shown in Fig. \ref{fig:absdrop}(a) do not reveal direct quantitative information about $\tilde \gamma$. However, the branching ratio can be extracted using an rms-coherent coupling $\langle \Omega^2_H(t) \rangle$ given in (\ref{eq:omegah_rms})

\begin{eqnarray}
\label{eq:branching}
\eta = \frac{\tilde \gamma}{\tilde \Gamma + \tilde \gamma} = \frac{\langle \Omega^2_H(t) \rangle}{\omega_z^2} =  \frac{B_\textrm{nuc}^2}{2 B_z^2}
\end{eqnarray}
$\eta$ is equivalent to the probability that the system decays via the $\tilde \gamma$-channel when excited into a trion state. 

To determine $\eta$ we applied a large square-wave modulation (Amplitude 80mV peak-peak) at different frequencies to the gate which in every cycle first loaded another electron of opposite spin into the QD forming a singlet together with the QD electron. Then one of the electrons was forced to leave, and as the tunneling probability for each of the two electrons is equal the remaining QD spin was fully randomized. The advantage of this technique which we will refer to as  \textit{electron cycling} leads to enforced spin relaxation at a known rate $\kappa_\textrm{exp}$ (also refer to equation (\ref{eq:kappasum})). In the case $\kappa \approx \kappa_\textrm{exp} \gg R_{1 \rightarrow 2}$ i.e. enforced spin relaxation rate exceeds the optical back-pumping rate $R_{2 \rightarrow 1}$ and $\widetilde{\gamma}$ can be determined by a fit using equations (\ref{eq:pop1122largek},\ref{eq:pop22largek}). 

\begin{figure}[h]
	\includegraphics[width=0.45\textwidth]{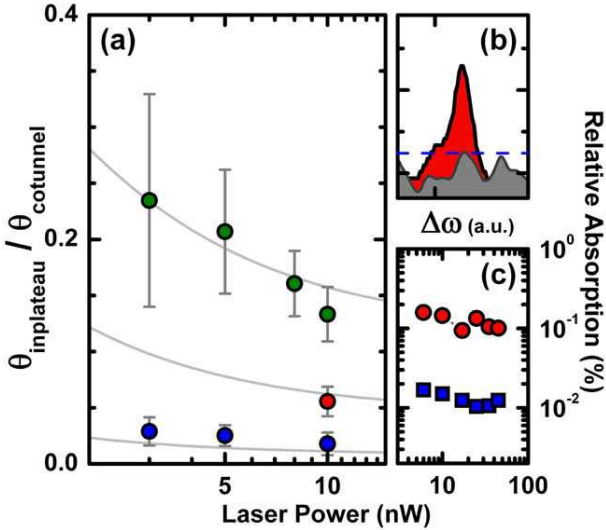}
	\caption{(Color online) (a) Electron recycling measurements. Peak absorption in the plateau center normalized to peak absorption in the co-tunneling regime is plotted as a function of laser power at a constant $B_z$=300mT and three different $\kappa = \kappa_\textrm{exp}$=54 kHz (upper, green points), 19kHz (middle, red points) and 3kHz (lower, blue points). The fits indicated by the solid grey lines have been obtained using equation (\ref{eq:pop22largek}) yielding $\gamma_\textrm{tot}^{-1}$= 0.63 $\mu$s. (b) A check experiment: Two laser scans at $\kappa = \kappa_\textrm{exp}$=54 kHz with in-and-out of plateau modulation (showing peak) and with in-plateau modulation, demonstrating that indeed controlled spin relaxation is realized. The noise level is indicated by the dashed blue line. (c) Intensity dependence of relative absorption at $B_z$=0T (red circles) and $B_z$=100 mT (blue squares). There is essentially no dependence on laser power confirming the theoretical model which gives (\ref{eq:rho22steadystate}). } 
	\label{fig:ecycling}
\end{figure}

Fig. \ref{fig:ecycling}(a) shows absorption normalized to on-resonance absorption in the co-tunneling regime; the data was obtained with electron cycling for different modulation frequencies and laser powers at a fixed external magnetic field $B_z$ = 300 mT. The upper, green points correspond to $\kappa$=54.3kHz, the middle, red point to $\kappa$ =19.3 kHz, and the lower, blue points to $\kappa$=3.3 kHz. 

Using (\ref{eq:thcalibration}), (\ref{eq:deftheta2}) and (\ref{eq:pop22largek}) the absorption ratio shown in the Figure can be written as

\begin{eqnarray}
\frac{\theta_\textrm{inplateau}}{\theta_\textrm{cotunnel}}= \frac{\rho_{22,\textrm{inplateau}}}{\rho_{22,\textrm{cotunnel}}} = \frac{2}{2 + \zeta}
\end{eqnarray}
The grey lines are fits using this expression. Best match with the data can be obtained with a total OSP rate $\gamma_\textrm{tot}$(300mT)=$\widetilde \gamma(300\textrm{mT}) + \gamma_{hm}$= 1.6$\mu$s$^{-1}$ \footnote{Due to our square-wave modulation scheme inducing a non-exponential spin decay the conversion from the known modulation frequency to $\kappa_\textrm{exp}$ rate does not need to include a factor of $2 \pi$. This has been confirmed by comparing the result of a numerical simulation using square wave-shaped spin relaxation to the outcome of a simple rate equation model including exponential decay of spin.}, where the two contributions stem from nuclear spins and hole mixing, respectively. The hole mixing-induced contribution can be independently determined from high-magnetic field measurements to be $\gamma_{hm}=2\mu$s. Using the branching ratio (\ref{eq:branching}) with $\widetilde \Gamma^{-1}$ = 0.8ns we then solve for the rms-nuclear magnetic field and obtain $B_\textrm{nuc}= 15 \textrm{mT}$ which is in good agreement with (\ref{eq:bnexperimental}).

Fig. \ref{fig:ecycling}(b) shows a measurement that demonstrates the difference between electron cycling (large amplitude modulation) and in-plateau (small amplitude) modulation: For in-plateau modulation we do not observe absorption (the noise level is marked by the horizontal dashed line). In contrast, when large amplitude modulation is applied absorption is partially recovered due to forced spin relaxation at a controlled rate $\kappa_\textrm{exp}$, as shown by the red peak in the figure. Fig.\ref{fig:ecycling}(c) was obtained in the plateau center without electron cycling technique, showing on-resonance absorption as a function of incident laser power. At 100 mT (blue squares) relative absorption is one order of magnitude weaker than at 0 mT (red circles) due to OSP. In both cases absorption exhibits weak dependence on laser power. This dependence arises from being in the vicinity of laser power required for saturation, hence a deviation from the assumptions of \eqref{eq:rho22steadystate}.

\subsection{\label{ssec:peakshift}Peak shift in the plateau center}

\begin{figure}[h]
	\includegraphics[angle = 0, width = 7.5cm]{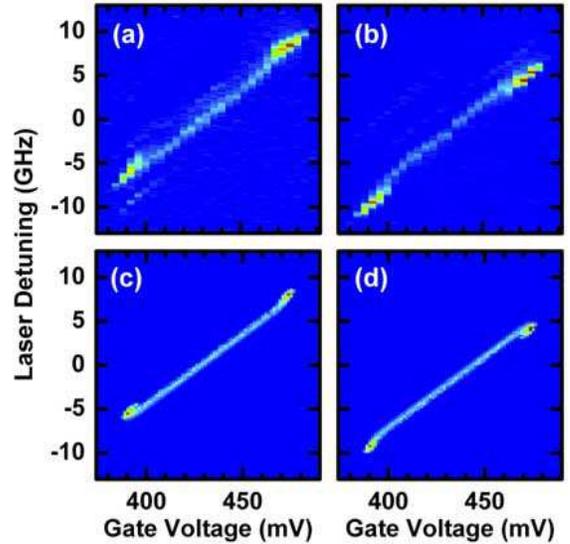}
	\caption{Gray-scale plot of absorption strength as a function of laser detuning and gate voltage. External magnetic field of $B_z$=150mT is applied. The linear gate-voltage dependence is due to the quantum-confined Stark effect. (a) and (c) show the absorption in the vicinity of the blue optical transition, (b) and (d) show the red transition. (a) and (b) are experimental data showing a spectral shift of the weak absorption peak in the plateau center compared to the strong co-tunneling regimes at the plateau edges. The shift is directed to the lower energies for the blue transition, and towards higher energies for the red transition. This feature can be reproduced in a numerical simulation including a randomly fluctuating Overhauser field as shown in (c) and (d) using the parameters $\Omega$ = 0.6 $\Gamma$, $\Gamma^{-1}$=0.8ns, $B_\textrm{nuc}$=15mT, $\gamma_{hm}^{-1}$=2$\mu$s.}
	\label{fig:lineshift}
\end{figure}

Fig. \ref{fig:lineshift} shows laser scans obtained at $B_z$=150mT on the blue (a) and red (b) Zeeman transition throughout the whole single-electron plateau. Absorption strength is grayscale-coded. The line tilt is due to the quantum confined Stark effect and pixelization is due to experimentally limited voltage resolution. At gate voltages 395 mV and 480 mV the two co-tunneling regimes show strong absorption when spin relaxation is fast due to charge reservoir coupling (for details refer to \ref{sec:cotunnel}). In the plateau center hyperfine interaction dominates and leads to spin pumping and drop of absorption as already discussed. Here, we further observe a shift of the spectral position of the absorption peak in the pleateau center as compared to the co-tunneling regime. This shift is directed to the red (blue) for the blue (red) Zeeman transition. This resembles effects one might expect for dynamical nuclear spin polarization (DNSP) \cite{LaiPRL06,EblePRB06}; these effects can however be excluded \footnote{As a Gedanken experiment we assume transfer of polarization from light to nuclei via the electron spin under a static positive magnetic field. Then, due to negative electron g-factor the spin-down state has higher energy than spin-up state. The laser is resonant with the red (first case) or the blue (second case) Zeeman transition. In the first case electron spin is pumped to the spin-up state which leads to nuclear spin-up polarization after an electron-nuclei flip-flop interaction. As the hyperfine constant $A$ is positive the energy of the electron spin-down state is consequently lowered while the trion states remain untouched, leading to a blue-shift of the red Zeeman optical transition which is being observed. The same happens in the second case where spin-down nuclear polarization is created and the energy of the spin-up state is lowered, again leading to a blue-shift of the observed transition. This scenario is clearly opposite of what we observe, thus ruling out the presence of an efficient DNSP.}. Our numerical simulation is able to reproduce this behaviour without taking into account DNSP as shown in Fig. \ref{fig:lineshift} (c) and (d). The lineshift in the plateau center is $\pm$0.9GHz for both red and blue transition which is close to the electronic Zeeman splitting at 150 mT, $E_{Z,e} \approx$ 1.3 GHz. At the co-tunneling edges $\kappa$ is large and maximum of absorption is observed when the laser is exactly on resonance with the transition. When $\kappa$ is small as it is the case in the plateau center absorption of a strictly resonant laser is suppressed due to spin pumping depending on the external magnetic field. When the laser frequency is moved towards the center between the strong $\widetilde\Gamma$ and the weak $\widetilde\gamma$ transitions i.e. the spectral detuning with respect to the $\widetilde\gamma$ transition is reduced, the back-pumping at rate $R_{1 \rightarrow 2}$ becomes more efficient and maximum of absorption will be reached for a spectral detuning that fulfils the condition $R_{1 \rightarrow 2} = R_{2 \rightarrow 1}$. As a consequence both transitions contribute to absorption which leads to a shift of the absorption maximum towards the weak $\widetilde \gamma$ transition i.e. a blueshift when the red line is observed and vice versa.

\subsection{\label{ssec:peakbroaden}Peak broadening at plateau edges}

\begin{figure}[h]
	\includegraphics[angle = 0, width = 7.5cm]{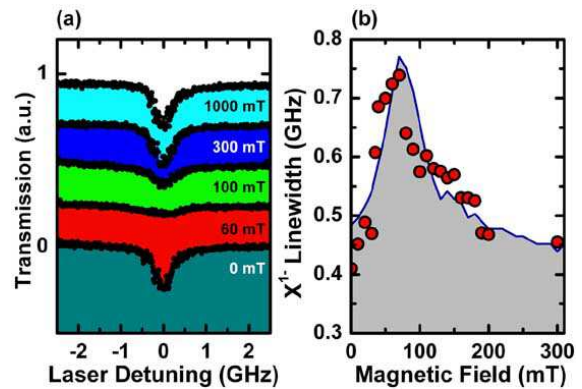}
	\caption{(Color online) (a) Trion transition laser scans for five different magnetic fields. The gate voltage was in the co-tunneling regime (see section \ref{sec:cotunnel}). (b) Measured linewidth obtained from laser scans as a function of magnetic field indicated by the red circles. A broadening occurs at $B_z \approx$75 mT; at larger fields the linewidth almost recovers back its original value of 450 MHz at 0 Tesla. The solid lines are obtained via numerical simulation with parameters $\Omega$ = 0.6$\Gamma$, $B_\textrm{nuc}$=15mT, $\Gamma^{-1}$=0.8ns, $\gamma_{hm}^{-1}=2.0\mu$s, $\kappa^{-1}$=2.5$\mu$s.}
	\label{fig:lwbroaden}
\end{figure}

Fig. \ref{fig:lwbroaden}(a) shows example laser scans for different magnetic fields $B_z$ ranging from 0 to 1 Tesla, obtained in the co-tunneling regime where spin relaxation is fast. The scans have been laterally shifted in order to eliminate the Zeeman shift. In (b) the measured linewidths are plotted as a function of magnetic field (red circles) along with a calculated curve (solid line). A broadening to almost double the zero-field linewidth appears at magnetic fields between 60 mT and 80 mT; at higher fields linewidth becomes as narrow as in the case $B_z=0$. 

The physical reason for the observed broadening is very similar to that described in the previous paragraph \ref{ssec:peakshift}: Both $\widetilde \gamma$ and $\widetilde\Gamma$ transitions contribute in a non-negligible way to absorption and maximum is observed when $R_{1 \rightarrow 2} = R_{2 \rightarrow 1}$ condition is fulfilled. Consistently the linewidth increases as much as the electronic Zeeman splitting initially but drops at magnetic fields where $\widetilde \Gamma(B) \gg \widetilde\gamma(B)$ and a single transition is established. In contrast to section \ref{ssec:peakshift} $\kappa$ is large here due to co-tunneling and annihilates the absorption drop caused by spin pumping hence making the transition visible at all magnetic fields. The solid lines are calculated curves for a range of $\varepsilon$ values using a randomly fluctuating Overhauser field with $B_\textrm{nuc}$ = 15mT, well reproducing this feature. We note that in order to put as many constraints as possible on the choice of simulation parameters we have used the maximum co-tunneling-induced spin relaxation rate $\kappa=\kappa_\textrm{cotunnel}= 0.4 \mu s^{-1}$ as obtained from the data shown in  Fig.\ref{fig:ctexp}(b). The effect of $\varepsilon$ on the simulation is clearly negligible, advocating that the dominant OSP mechanism is hyperfine interaction.


\subsection{Coupling to electron spin reservoir}

\begin{figure}[h]
	\includegraphics[angle = 0, width = 8.2cm]{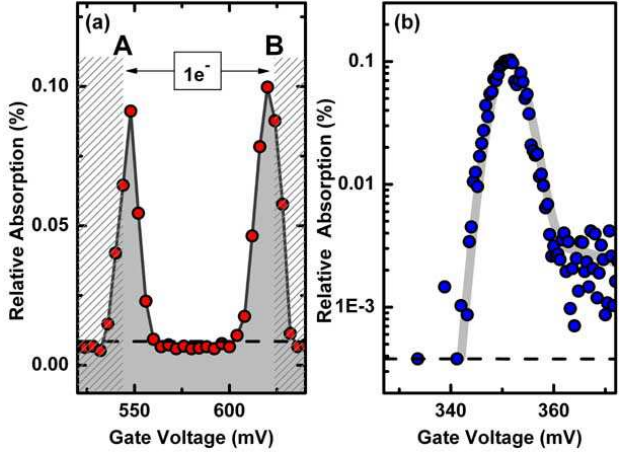}
	\caption{(Color online) (a) Example voltage coarse scan across the entire single-electron plateau at $B_z$=300mT. Per voltage step a laser scan is performed and the observed absorption maximum is plotted as a function of gate voltage.	Due to strong hyperfine-induced OSP and weak co-tunneling rate absorption in the plateau center is suppressed; however when approaching the single-electron plateau edges absorption is recovered due to highly nonlinear dependence of co-tunneling on gate voltage leading to fast spin flips. Outside the voltage plateau i.e. left of point A or right of point B absorption is suppressed because the QD then becomes either empty or doubly-charged which shifts the optical transition energies out of our spectral observation window of 30GHz. The solid line is a guide to the eye. (b) shows a voltage fine scan of the left co-tunneling regime obtained on another QD than in (a). The solid line is a numerical calculation using $\Gamma^{-1}$ = 0.8ns, $\Omega$=0.6 $\Gamma$, $B_\textrm{nuc}=$15mT, $\Gamma_\textrm{tunnel}^{-1}$=20ns, $\gamma_{hm}^{-1}=2\mu$s and a Zeeman splitting of $E_{Z,e}=10\mu$eV. The voltage-FWHM of the co-tunneling peak is 10mV.}
\label{fig:ctexp}
\end{figure}

We have performed laser scans as a function of gate voltage througout the whole single-electron plateau as defined in section \ref{sec:cotunnel}. The measured on-resonance absorption signal for each laser scan is plotted in Fig. \ref{fig:ctexp}(a). The data has been obtained at an external magnetic field of $B_z$=300 mT taking coarse voltage steps. 

At gate voltages lower than 540 mV and higher than 625 mV as marked by the shaded regions absorption drops below noise level indicated by the horizontal dashed line. At these voltages the QD either becomes empty (left of point A) or doubly charged (right of point B). Absorption then vanishes since in those cases the QD is not described by the trion level system any more; the optical transitions for these gate voltages are not observed within our scanning window of 30 GHz around the trion transitions. The unshaded part indicates the region where the QD contains a single electron and as it has been mentioned before the co-tunneling rate is maximum when gate voltage is at the crossover points A or B. Here, relaxation via co-tunneling is faster than the optical pumping rates $\kappa_\textrm{cotunnel} \gg R_{1 \rightarrow 2}, R_{2 \rightarrow 1}$ leading to thermalization of the electron spin and thus strong absorption. The scenario drastically changes when gate voltage is tuned to the center of the plateau. Here, co-tunneling rate $\kappa_\textrm{cotunnel}$ reaches its minimum where our numerical calculation predicts a drop of as much as five orders of magnitude (also see Fig. \ref{fig:cotcalc}) compared to the crossover points such that $\kappa_\textrm{cotunnel} \ll R_{2 \rightarrow 1}$. Consequently the occupation of the spin states is governed by OSP (equation (\ref{eq:rho22steadystate})) rather than Boltzmann factor meaning that the spin is predominantly in the dark state and vanishing absorption is observed. 


The semilogarithmic plot in Fig. \ref{fig:ctexp}(b) shows a voltage fine scan of the low voltage plateau edge around the A crossover point obtained at $B_z$=300mT. The gate voltage for point A is different from Fig. \ref{fig:ctexp}(a) as this data was taken on another QD. The observed absorption drops by half within a gate voltage detuning of $\pm$5mV from the maximum position. This data demonstrates the enormous gate-voltage dependence of this spin relaxation mechanism. The gray solid line is a best-fit numerical simulation using expression (\ref{eq:cotunnel}) as spin relaxation rate showing good accordance with the data. The co-tunneling rate at the peak as determined from the fit is $\kappa_\textrm{max}^{-1}$=2.5$\mu$s. The noise level is indicated by the dashed line; it deviates from the one shown in (a) due to different experimental settings such as lock-in time constants and filter slopes. Again on the left side of the peak the QD is empty, yielding vanishing absorption below the noise level. The gradual decrease of absorption is due to finite temperature. On the right side the spin pumping regime is located; here some weak absorption remains according to the occupation of the observed spin state, revealing the strength of spin pumping.

\subsection{Coupling to phonon reservoir}
\label{sec:phononexp}

\begin{figure}[h]
	\includegraphics[angle = 0, width = 8.6cm]{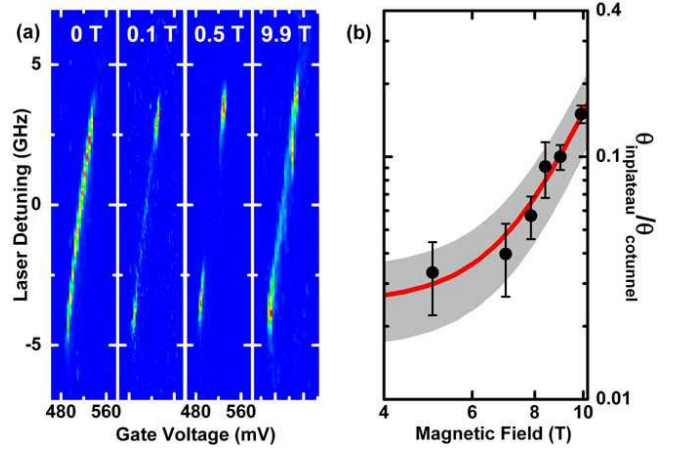}
	\caption{(Color online) (a) Plateau laser scans for 4 different magnetic fields: Absorption is plotted as a function of laser detuning and gate voltage. Zeeman effect has been eliminated by vertical shift of each single plot. The linear voltage dependence is due to quantum-confined Stark effect. Plateau center absorption drops at intermediate magnetic fields due to OSP but then recovers at high fields due to fast thermalization via phonon-SO interaction at 9.9 Tesla. At the plateau edges the QD can absorb at all magnetic fields due to fast $\kappa_\textrm{cotunnel}$.	(b) Black data points: Magnetic-field evolution of peak absorption in the plateau center normalized to peak absorption in the co-tunneling regime. The lower (upper) boundary of the grey region is obtained from a numerical simulation for $\gamma_{hm}^{-1}=1.2\mu$s ($\gamma_{hm}^{-1}=2.8\mu$s); the red line corresponds to $\gamma_{hm}^{-1}=2\mu$s. Best match with the data yields explicitly for the spin relaxation rate $\kappa_\textrm{phonon} = \alpha_0 B^5$ with $\alpha_0=0.031$ in units of [T$^{-5}$sec$^{-1}]$. Again, $B_\textrm{nuc}$=15mT.}
	\label{fig:sophonon}
\end{figure}

Based on the theoretical estimates of section \ref{sec:phonondecay} we now seek for signatures of SO-phonon induced spin relaxation at high magnetic fields. Fig.\ref{fig:sophonon}(a) shows 2D plots of color-coded absorption strength as a function of laser detuning and gate voltage for four different magnetic fields obtained for the red Zeeman transition. The scans cover the whole single-electron plateau; excitonic Zeeman shift has been eliminated by shifting the y-scale for each 2D graph separately. The linear dependence of the excitonic transition energy on gate voltage is due to the quantum confined stark shift. 

At 0 Tesla absorption is clearly visible throughout the whole plateau due to fast spin flips with the neighboring nuclear spins. When a small magnetic field (B=0.1T) is applied, absorption in the plateau center drops because of hyperfine-induced OSP as disucssed in the previous sections. Close to the plateau edges absorption still remains due to fast co-tunneling. At 0.5 Tesla increasing OSP leads to further drop of absorption. These absorption characteristics in the plateau center remain the same up to 5 Tesla, however absorption starts to come back at even higher fields: when the magnetic field is raised up to 9.9 Tesla a significant recovery of plateau-center absorption is observed. This effect cannot be explained by OSP which only causes monotonous decrease of absorption nor by co-tunneling which is negligible in the plateau center and hardly shows any magnetic field dependence. Owing to its $B^5$ dependence however, phonon-assisted spin relaxation is a good candidate to be responsible for the observed effects in the context of spin relaxation mechanisms. 

Fig.\ref{fig:sophonon}(b) shows the quantitative evolution of normalized absorption with magnetic field i.e. the ratio of absorption in the plateau center versus co-tunneling regime (black data points). Further, the solid red line along with the gray shaded region indicate calculated strength of absorption for $\gamma_{hm}^{-1}=2 \mu$s with an uncertainty of $\pm 0.8 \mu$s; the phonon-induced spin relaxation rate was $\kappa_\textrm{phonon} = \alpha_0 B^5$ with the coefficient $\alpha_0=0.031$ in units of [T$^{-5}$sec$^{-1}]$. Whereas $\kappa_\textrm{phonon}$ is strongly $B$-dependent, the hole mixing contribution $\gamma_{hm}$ has no $B$-field dependence within the magnetic field range considered here. Therefore these two mechanisms have distinguishable effects on Fig.\ref{fig:sophonon}(b) and thus can be identified independently. The good agreement with the experimental data strongly suggests that in this regime of electric and magnetic fields the dominant spin relaxation is indeed phonon-assisted. Further, within our uncertainty $\kappa_\textrm{phonon}$ matches well with the results that have been previously obtained on an ensemble of self-assembled InAs/GaAs QDs \cite{KroutvarN04}.

There are two fundamentally different mechanisms which employ holes to yield OSP: First, hole mixing of strength $\varepsilon$ leads to an admixture of the light hole states to the trion states as dicussed in section \ref{ssec:holemix}. Second, hole-spin \textit{relaxation} leads to an incoherent coupling of the trion states contributing to OSP. In \cite{BulaevPRL05} hole spin relaxation rate is predicted to be below $10^3$/sec and monotonically increase with magnetic field which suggests that it is not the main mechanism responsible for OSP. We therefore neglect hole spin-flips, further assuming that there is no other efficient hole spin flip mechanism at low magnetic fields. In the first mechanism, OSP is independent of magnetic field and the strength is equal to the hyperfine-induced OSP rate at $\sim$1 Tesla. At higher magnetic fields the hyperfine-induced OSP rate drops with $B^{-2}$, therefore hole mixing becomes the dominant OSP mechanism here.


\section{Full interaction map}
\label{sec:fullmap}

\begin{figure}[h]
	\includegraphics[angle = 0, width = 7.7cm]{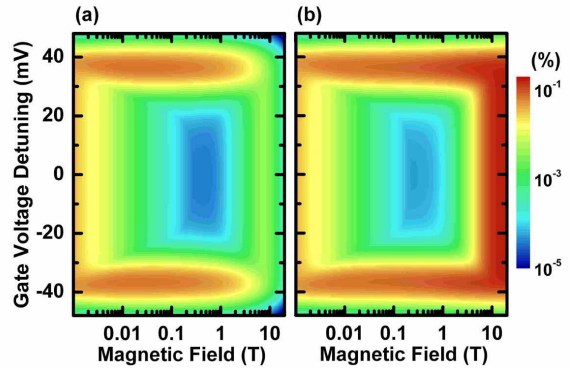}
	\caption{(Color online) Calculated absorption maxima for the whole single-electron plateau plotted as a function of magnetic and electric field. (a) shows the simulation for the probe laser in the vicinity of the red Zeeman transition, (b) similar but for the blue Zeeman transition. The borders of each plot show strong absorption due to  interactions with nuclear spins (left), charge reservoir (top/bottom) and phonon reservoir (right). At large magnetic fields spin polarization nearly reaches unity due to thermalization leading to vanishing absorption on the red transition (a), and enhanced absorption on the blue transition (b). In the center of the plots absorption and thus spin relaxation is suppressed by approximately five orders of magnitude. The parameters used in the simulation are: $\Gamma^{-1}$=0.8ns, $B_\textrm{nuc}$=15mT, $\gamma_{hm}^{-1}$=2$\mu$s, tunneling time $\Gamma_\textrm{tunnel}^{-1}$=20ns, $\kappa_\textrm{phonon}$ as given in section \ref{sec:phononexp}. }
	\label{fig:fullmap}
\end{figure}

In the previous sections the three dominant mechanisms acting on the confined spin have been identified separately along with the two mechanisms for OSP. In this final part we will present results of our numerical simulation based on parameters that have been measured or estimated before. The calculations have been performed within a parameter space approximately overlapping with the full scale of our experimental tuning ability of the static electric and magnetic fields. For details of the simulation we refer to Appendix \ref{sec:levelx}.

Fig. \ref{fig:fullmap} shows calculated maximum values of absorption for the red (a) and the blue (b) trion transition with a laser having the corresponding circular polarization. Absorption strength is color-coded in logarithmic scale as a function of gate voltage detuning and external magnetic field. 

All of the following points have been discussed in the previous chapters; here we mention them briefly as a key to the plots: the necessary conditions for observing strong absorption are either $\kappa \gg R_{1 \rightarrow 2},R_{2 \rightarrow 1}$ or $R_{1 \rightarrow 2} \sim R_{2 \rightarrow 1}$. Further, at large magnetic fields $B >8$ Tesla when $E_{Z,e}\sim kT$ the Boltzmann factor leads to a difference of the spin ground state occupations and thus a difference between absorption strength on the red and blue Zeeman transition. 

In the plot we distinguish three different regimes of strong absorption:

\begin{enumerate}
	\item \textit{Magnetic fields lower than the fluctuations of the hyperfine field (B $\lesssim$15 mTesla)}. Here, fast bidirectional OSP due to hyperfine-induced state mixing leads to strong absorption. 

	\item \textit{High magnetic fields ($>$5 Tesla)}. Here, $\kappa_\textrm{phonon}$ induces fast thermalization i.e. $\kappa_\textrm{phonon} \gg R_{1 \rightarrow 2},R_{2 \rightarrow 1}$. The spin ground state occupation is mainly determined by the Boltzmann factor leading to a lowering (increase) of absorption on the higher (lower) energy spin state occupation (a) ((b)).

	\item \textit{Large gate voltage detunings from the plateau center ($\pm$40mV)}. Here, co-tunneling ($\kappa_\textrm{cotunnel}$) is responsible for fast spin relaxation and appearance of absorption. 
\end{enumerate}

An intriguing feature that becomes apparent now is the blue \textit{island} in the center of the color-coded spin relaxation plot. It marks the regime where absorption (i.e. all reservoir interactions) is suppressed by five orders of magnitude, or in other words the localized spin becomes maximally isolated hence the frequently used concept of an \textit{artificial atom} is meaningful. Within the scope of quantum information processing this indicates the relevant regime of operation.


\section{\label{sec:conclusions}Summary and Conclusion}

We have investigated the dominant interactions of a confined electron spin in a single self-assembled QD by optical means and demonstrated the regimes where each reservoir coupling becomes important. For magnetic fields $B \lesssim 1$Tesla, the dominant contribution to OSP stems from the fluctuating hyperfine field mixing the electronic spin states and creating a weak channel for diagonal relaxation in the trion four-level picture. Exchange and phonon-induced spin-flip processes dominate over hyperfine-induced spin pumping and establish a thermal steady-state at large gate-voltage detunings and/or large external magnetic fields; in the plateau center at intermediate magnetic fields the situation is reversed and spin pumping dominates, strongly altering the state occupations away from thermal equilibrium values. Signatures of heavy-light hole mixing dominated spin cooling can be observed for fields $\gtrsim 5$Tesla.

From a quantum control perspective these results demonstrate that the quantum dynamics of a single confined spin can be significantly altered by externally controlled parameters such as electric and magnetic fields. A natural extension of this study would be the investigation of spin decoherence in a single QD using similar optical techniques. These measurements would require more advanced schemes such as electromagnetically-induced transparency (EIT). Further, knowledge gained on single-electron spin dynamics can be utilized in the resonant optical study of more complex systems such as coupled QDs or QDs with a single excess heavy-hole.

\begin{acknowledgments}

We thank Hakan T\"ureci, Alex H\"ogele, Tunc Yilmaz, Jeroen Elzerman, and Nick Vamivakas for useful discussions.
This work is supported by NCCR Quantum Photonics. 
J.D. and M.A. would like to thank J. Cash for technical assistance.

\end{acknowledgments}

\appendix


\section{\label{app:sample}Sample and experimental techniques}

\begin{figure}[h]
	\includegraphics[scale=0.41]{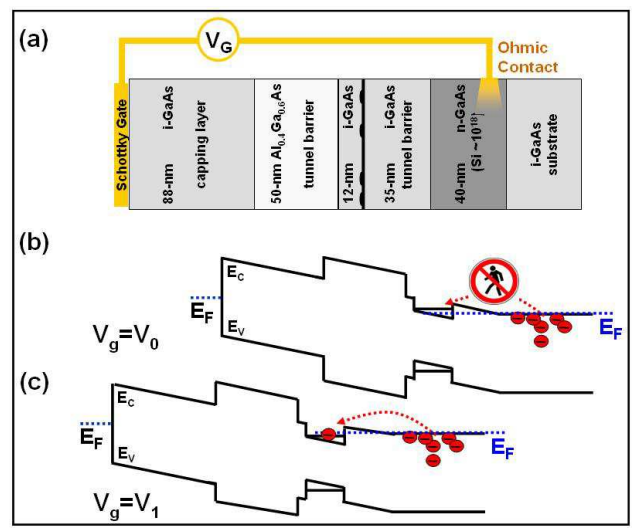}
	\caption{(a) Layers of the Schottky-type heterostructure. (b) Corresponding band structure. At gate voltage $V_0$ the lowest QD conduction band level lies above the Fermi energy and the QD is empty. (c) At gate voltage $V_1 > V_0$ the lowest QD level is below the Fermi energy and therefore populated with one electron. A second electron cannot enter due to required charging energy (\textit{Coulomb blockade}). }
		\label{fig:schottky}
\end{figure}

Our InAs/GaAs quantum dots (QDs) are grown by molecular beam epitaxy in Stranski-Krastanow mode leading to lens-shaped dots of average size 25nm$\times$25nm$\times$5nm; QD light emission is blue-shifted by partially covered islands (PCI) technique. A 35-nm GaAs tunneling barrier separates the QDs from a charge reservoir formed by a heavily doped n-GaAs layer which forms the back contact. Above the QDs there is a 12 nm thick GaAs cap and a 50 nm-thick Al$_{0.4}$Ga$_{0.6}$As blocking layer which prevents the holes from coupling to the continuum states within the 88-nm capping layer \cite{SeidlPRB05}. Bias voltage between the back contact and a semi-transparent 5 nm-Ti-Schottky window determines the electric field in the structure and allows us to load a single conduction-band electron into the QD. 
\begin{figure}[h]
	\includegraphics[scale=0.4]{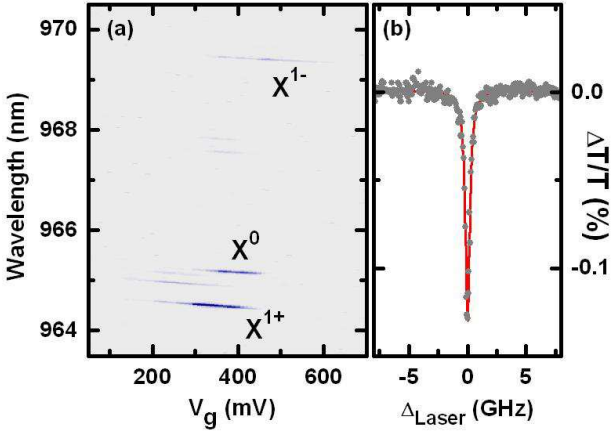}
	\caption{(a) Example gatesweep. This plot has been obtained by increasing the gate voltage step by step and for each step taking a single-QD photoluminescence spectrum. The three strongest emission lines are identified as $X^0$, $X^{1-}$ and tentatively $X^{1+}$ which result from s-shell electron-hole recombination from differently charged excitonic complexes. Discrete and characteristic PL energies, different for every exciton complex, are induced by Coulomb interaction. Small continuous PL energy shift is due to quantum confined stark shift.	(b) Example differential transmission laser scan.	When a laser is scanned over a QD transition light is resonantly Rayleigh-scattered. In DT experiment this Rayleigh-scattered light interferes with the laser background resulting in a dip in the transmitted light. The FWHM of the Lorentzian fit indicated by the solid red line is 460MHz.}
		\label{fig:pldt}
\end{figure}

All experiments described here are carried out with a confocal microscopy setup immersed in a liquid helium bath cryostat at a temperature of 4.2 Kelvin. The numerical aperture of the microscope is $0.68$ resulting in a diffraction limited spot size of $\sim 1\mu m$. Area density of QDs in our sample is low enough to have $\leq$ 3 dots in the focal spot simultaneously. Different QDs can then easily be spectrally separated by their inhomogeneous broadening. A magnetic field of up to 10 Tesla along the z-axis can be applied by a superconducting magnet. Piezo-electric nanopositioners allow us to move the sample in XYZ-space. Transmitted light is collected and sent to a circular polarization analyzer which distributes the light to two photodetectors, similar to that of Ref.\cite{AtatureS06}.

The initial step of our experiment is a gatesweep i.e. a PL measurement as a function of gate voltage. For this we send in a laser exciting electrons and holes in the bulk GaAs at an energy of $\approx$1.6 eV. After filtering out the pump light the resulting QD luminescence is sent to a grating spectrometer with a resolution $\sim 20\mu$eV and detected by a liquid-nitrogen cooled CCD. 

Hereafter the differently charged excitonic complexes can be identified by their characteristic emission energy and voltage dependence profile \cite{WarburtonN00}. From then on we only apply resonant excitation of the QD single electron ground state transitions by using a differential transmission (DT) technique \cite{AlenAPL06,AlenAPL03,HogelePRL04,HogelePE04}. In order to obtain a spectrum we sweep a single-mode Ti:Sa laser over the QD transition and record the intensity of the transmitted light. A QD resonance is observed as a dip on top of the laser background. The resolution of this technique is only limited by the laser linewidth i.e. $\Delta \nu_{Laser} <$1 MHz.

%

\section{\label{ssec:obe}Optical Bloch equations for the three-level system}

The optical Bloch equations are derived from the master equation \eqref{eq:mastereq}. Including rotating-wave approximation and taking the limit of $E_z \ll kT$ which eliminates the Boltzmann factors, the optical Bloch equations read using the basis states introduced in \eqref{eq:oldbasis}

\begin{eqnarray*}
\label{system1}
 \frac{\textrm d}{\textrm d t}\;\rho_{11} &=& \textrm i \Omega_H(\rho_{12}-\rho_{21})+\gamma_{hm} \rho_{33} -\kappa(\rho_{11}-\rho_{22}) \\ \nonumber
\label{system3}
  \frac{\textrm d}{\textrm d t}\;\rho_{22} &=& \textrm i\frac{\Omega_R}{2}(\rho'_{23}- \rho'_{32}) + \textrm i \Omega_H(\rho_{21}-\rho_{12})  \\ \nonumber
 && + \Gamma\rho_{33} + \kappa(\rho_{11}-\rho_{22}) \\
\label{system2}
  \frac{\textrm d}{\textrm d t}\;\rho_{33} &=& \textrm i\frac{\Omega_R}{2}(\rho'_{32}-\rho'_{23})-(\Gamma+\gamma_{hm})\rho_{33} \\
  \frac{\textrm d}{\textrm d t}\;\rho_{12} &=& \textrm i\frac{\Omega_R}{2} \rho'_{13} + \textrm i \Omega_H(\rho_{11}-\rho_{22}) - \kappa\rho_{12} \\
 \frac{\textrm d}{\textrm d t}\; \rho'_{13} &=&  \textrm i\frac{\Omega_R}{2}\rho_{12} - \textrm i \Omega_H \rho'_{23}  \\
\nonumber 
&&+ (-\frac{\Gamma+\gamma_{hm}+\kappa}{2}-\textrm i\delta\omega)\rho'_{13} \\
  \frac{\textrm d}{\textrm d t}\;\rho'_{23} &=& \textrm i\frac{\Omega_R}{2}(\rho_{22}-\rho_{33}) -  \textrm i \Omega_H \rho'_{13} \\
\nonumber  
&&  + (-\frac{\Gamma+\gamma_{hm}+\kappa}{2}-\textrm i \delta\omega)\rho'_{23}
\end{eqnarray*}
with
\begin{eqnarray*}
  \rho_{13} &=& \rho'_{13} \: e^{\textrm i \omega_L t } \\
  \rho_{23} &=& \rho'_{23} \: e^{\textrm i \omega_L t }
\end{eqnarray*}
We have $\rho'_{31} = \rho'^{*}_{13}$, $\rho'_{21} = \rho'^{*}_{12}$, $\rho'_{32} = \rho'^{*}_{23}$, and $\rho_{11} +\rho_{22} +\rho_{33}= 1$.


\section{\label{ssec:sytrafo}Dressed-state Transformation}

The transformation used to diagonalize the coupling to the quasi-static nuclear (Overhauser) field can be written as

\begin{eqnarray}
\widetilde H = S \hat H S^\dagger \\ \nonumber
\widetilde \rho = S \rho S^\dagger
\end{eqnarray}
with $\phi = \Omega_H / \omega_z$ and $S^\dagger S =  \mathbb{I}$. We assume $\phi \ll 1$. The new basis (\ref{eq:newbasis}) determines the transformation matrix. When taking only first order terms

\begin{eqnarray}
S = \begin{pmatrix}
1 & -\phi & 0 \cr
\phi & 1 & 0 \cr
0 & 0 & 1 \cr
\end{pmatrix} 
\end{eqnarray}
The spontaneous emission terms then yield 
 
\begin{eqnarray}
S  L_{\textrm{relaxation},\Gamma} S^\dagger = \frac{ \Gamma}{2} ( 2 S \sigma_{23} S^\dagger ~ \widetilde \rho ~ S \sigma_{32} S^\dagger \\ \nonumber
- S \sigma_{33} S^\dagger \widetilde\rho -\widetilde \rho S \sigma_{33} S^\dagger)
\end{eqnarray}
For the new projection operator $S \sigma_{23} S^\dagger$ we obtain

\begin{eqnarray}
S \sigma_{23} S^\dagger = \phi \sigma_{\tilde 1 \tilde 3} + \sigma_{\tilde 2 \tilde 3}
\end{eqnarray}
and the conjugate relation. Here, $\sigma_{\tilde i \tilde j}=  |\widetilde i\rangle \langle \widetilde j|$. Using this with the previous relation, we obtain

\begin{eqnarray}
S L_{\textrm{relaxation},\Gamma} S^\dagger &=& \frac{ \widetilde \gamma}{2} ( 2 \sigma_{\tilde 1 \tilde 3} ~ \widetilde \rho ~  \sigma_{\tilde 3 \tilde 1}  - \sigma_{\tilde 3 \tilde 3}\widetilde\rho -\widetilde \rho \sigma_{\tilde 3 \tilde 3})\\ \nonumber
&&+ \frac{\widetilde \Gamma}{2} ( 2 \sigma_{ \tilde 2 \tilde 3} \widetilde\rho \sigma_{\tilde 3 \tilde 2} - \sigma_{\tilde 3 \tilde 3}\widetilde\rho -\widetilde \rho \sigma_{\tilde 3 \tilde 3}) \\ \nonumber
&&-2 \phi \frac{\Gamma}{2} (\sigma_{\tilde 2 \tilde 3} \widetilde\rho \sigma_{\tilde 3 \tilde 1} +\sigma_{\tilde 1 \tilde 3}\widetilde \rho \sigma_{\tilde 3 \tilde 2})
\end{eqnarray}
where $\widetilde \gamma = \phi^2 \Gamma$ and $\widetilde \Gamma = \Gamma$. $\widetilde \sigma_{ij} = |\widetilde i\rangle \langle \widetilde j|$ is the projection operator acting on $\widetilde \rho$. The first, $\widetilde \gamma$ term corresponds to relaxation via a weak optical transition induced by the hyperfine field, allowing for spin-flip Raman events and the second, $\widetilde \Gamma$ term, describes relaxation via the strong optical trion transition. 

The last term describes coherence induced by the spontaneous relaxation into a superposition of dressed-basis ground states at a rate proportional to the occupation of the excited state $\widetilde \rho_{33}$. When multiplying with $\langle 2|$ from the left and $|1 \rangle$ from the right we obtain
\begin{eqnarray}
\frac{\textrm d}{\textrm d t} \widetilde \rho_{21} = -2 \phi \Gamma \widetilde \rho_{33}
\end{eqnarray}
The same relation is obtained for $\widetilde \rho_{21}$ when multiplying with $\langle 1|$ and $|2 \rangle$ respectively. 

The transformed $\kappa$ terms keep the Lindblad form and we obtain for $E_{Z,e} \ll kT$
\begin{eqnarray}
\widetilde L_{\textrm{relaxation},\kappa}= \frac{\widetilde \kappa}{2}  \left[ ( 2 \widetilde N_{21} \widetilde\rho \widetilde N_{12}- \widetilde M_{11}\widetilde\rho -\widetilde \rho \widetilde M_{11}) \right. \\ \nonumber \left.
+ ( 2 \widetilde N_{12} \widetilde\rho \widetilde N_{21}- \widetilde M_{22}\widetilde\rho -\widetilde \rho \widetilde M_{22})
\right] 
\end{eqnarray}
with $\widetilde N_{21}= S \sigma_{21} S^\dagger, \widetilde N_{12}=\widetilde N_{21}^\dagger, \widetilde M_{ii}= S \sigma_{ii} S^\dagger$ and $\widetilde \kappa =  \kappa$.

%

\section{Numerical Studies}
\label{sec:levelx}
The derived formalism considers a static randomly oriented nuclear field. Within a measurement time $B_N$ changes $\sim$100 times. In order to calculate measurable quantities such as linewidths and peak heights as functions of electric and magnetic field we have thus performed numerical simulations: For a given set of parameters the steady-state solutions of the optical Bloch equations as given in Appendix \ref{ssec:obe} are numerically evaluated, in particular $\textrm{Im}(\rho_{23}(\infty))$ is then linked to the absorption (details in section \ref{sec:spectrosc}). A fluctuating hyperfine field is implemented by pulling three random numbers $B_{N,i}$ following (\ref{eq:bni_gauss}). From $B_{N,xy}$, using (\ref{eq:bnxy}), state-mixing strength $\Omega_H$ (\ref{eq:defomegah}) and pure Zeeman splitting $\omega_z$ are calculated before evaluating the density matrix steady-state. This procedure is repeated in order to average over  $\sim$100 random settings of the hyperfine field. In the cases the simulation could not be performed throughout the whole parameter space, we confirmed in key regimes that results agree well with that of a static Overhauser field with equal magnitude in $x,y,z$:
From (\ref{eq:bni_gauss}) and (\ref{eq:defomegah}) we obtain for the rms-value of $\Omega_H(t)$
\begin{eqnarray}
\label{eq:omegah_rms}
\langle \Omega^2_H(t) \rangle =  \left(\frac{g_e \mu_B }{\hbar} \right)^2 \frac{ \langle  B^2_{xy}(t)\rangle}{4} =  \left(\frac{g_e \mu_B }{\hbar} \right)^2 \frac{B_\textrm{nuc}^2}{2}
\end{eqnarray}
Here, the assumption for the observed absorption $\Theta$ to be made is
\begin{eqnarray}
\langle ~\Theta \left( B^2_{N,xy}(t) \right)~ \rangle  &\approx& \Theta \left(~ \langle B^2_{N,xy}(t) \rangle ~\right) 
\end{eqnarray}
i.e., the averaging over the absorption strength for different settings of the hyperfine field approximately equals the strength of absorption for the average field magnitude, equal in $x,y,z$.


\bibliographystyle{pf}

\begin{thebibliography}{10}

\bibitem{LossPRA98}
D.~Loss and D.~P. DiVincenzo,
\newblock Phys. Rev. A {\bf 57}, 120 (1998).

\bibitem{ImamogluPRL99}
A.~Imamoglu, D.~D. Awschalom, G.~Burkard, D.~P. DiVincenzo, D.~Loss,
  M.~Sherwin, and A.~Small,
\newblock Physical Review Letters {\bf 83}, 4204 (1999).

\bibitem{CalarcoPRA03}
T.~Calarco, A.~Datta, P.~Fedichev, E.~Pazy, and P.~Zoller,
\newblock Physical Review A {\bf 68}, 012310 (2003).

\bibitem{AtatureS06}
M.~Atat\"ure, J.~Dreiser, A.~Badolato, A.~H\"ogele, K.~Karrai, and A.~Imamoglu,
\newblock Science {\bf 312}, 551 (2006).

\bibitem{ElzermanN04}
J.~Elzerman, R.~Hanson, L.~Van~Beveren, B.~Witkamp, L.~Vandersypen, and
  A.~Kouwenhoven,
\newblock Nature {\bf 430}, 431 (2004).

\bibitem{Amasha06}
S.~Amasha, K.~MacLean, I.~Radu, D.~M. Zumbuhl, M.~A. Kastner, M.~P. Hanson, and
  A.~C. Gossard,
\newblock arXiv:cond-mat/0607110v1 [cond-mat.mes-hall]  (2006).

\bibitem{JohnsonN05}
A.~Johnson, J.~Petta, J.~Taylor, A.~Yacoby, M.~Lukin, C.~Marcus, M.~Hanson, and
  A.~Gossard,
\newblock Nature {\bf 435}, 925 (2005).

\bibitem{PettaS05}
J.~R. Petta, A.~C. Johnson, J.~M. Taylor, E.~A. Laird, A.~Yacoby, M.~D. Lukin,
  C.~M. Marcus, M.~P. Hanson, and A.~C. Gossard,
\newblock Science {\bf 309}, 2180 (2005).

\bibitem{KoppensS05}
F.~H.~L. Koppens, J.~A. Folk, J.~M. Elzerman, R.~Hanson, L.~H.~W. van Beveren,
  I.~T. Vink, H.~P. Tranitz, W.~Wegscheider, L.~P. Kouwenhoven, and L.~M.~K.
  Vandersypen,
\newblock Science {\bf 309}, 1346 (2005).

\bibitem{KoppensN06}
F.~Koppens, C.~Buizert, K.~Tielrooij, I.~Vink, K.~Nowack, T.~Meunier,
  L.~Kouwenhoven, and A.~Vandersypen,
\newblock Nature {\bf 442}, 766 (2006).

\bibitem{KroutvarN04}
M.~Kroutvar, Y.~Ducommun, D.~Heiss, M.~Bichler, D.~Schuh, G.~Abstreiter, and
  J.~Finley,
\newblock Nature {\bf 432}, 81 (2004).

\bibitem{MerkulovPRB02}
I.~A. Merkulov, A.~L. Efros, and M.~Rosen,
\newblock Phys. Rev. B {\bf 65}, 205309 (2002).

\bibitem{ErlingssonPRB02}
S.~I. Erlingsson and Y.~V. Nazarov,
\newblock Phys. Rev. B {\bf 66}, 155327 (2002).

\bibitem{Maletinskya07}
P.~Maletinsky, A.~Badolato, and A.~Imamoglu,
\newblock arXiv:0704.3684v1 [physics.optics]  (2007).

\bibitem{Taylor2006}
J.~M. Taylor, J.~R. Petta, A.~C. Johnson, A.~Yacoby, C.~M. Marcus, and M.~D.
  Lukin,
\newblock arXiv:cond-mat/0602470v2 [cond-mat.mes-hall]  (2006).

\bibitem{MaletinskyPRB07}
P.~Maletinsky, C.~W. Lai, A.~Badolato, and A.~Imamoglu,
\newblock Physical Review B {\bf 75}, 035409 (2007).

\bibitem{AverinPRL90}
D.~V. Averin and Y.~V. Nazarov,
\newblock Phys. Rev. Lett. {\bf 65}, 2446 (1990).

\bibitem{Goldhaber-GordonN98}
D.~Goldhaber-Gordon, H.~Shtrikman, D.~Mahalu, D.~Abusch-Magder, U.~Meirav, and
  M.~A. Kastner,
\newblock Nature {\bf 391}, 156 (1998).

\bibitem{CronenwettS98}
S.~M. Cronenwett, T.~H. Oosterkamp, and L.~P. Kouwenhoven,
\newblock Science {\bf 281}, 540 (1998).

\bibitem{GovorovPRB03}
A.~O. Govorov, K.~Karrai, and R.~J. Warburton,
\newblock Phys. Rev. B {\bf 67}, 241307 (2003).

\bibitem{HelmesPRB05}
R.~W. Helmes, M.~Sindel, L.~Borda, and J.~von Delft,
\newblock Physical Review B {\bf 72}, 125301 (2005).

\bibitem{SeidlPRB05}
S.~Seidl, M.~Kroner, P.~A. Dalgarno, A.~H\"ogele, J.~M. Smith, M.~Ediger, B.~D.
  Gerardot, J.~M. Garcia, P.~M. Petroff, K.~Karrai, and R.~J. Warburton,
\newblock Physical Review B {\bf 72}, 195339 (2005).

\bibitem{SmithPRL05}
J.~M. Smith, P.~A. Dalgarno, R.~J. Warburton, A.~O. Govorov, K.~Karrai, B.~D.
  Gerardot, and P.~M. Petroff,
\newblock Physical Review Letters {\bf 94}, 197402 (2005).

\bibitem{D'yakonovSP-J86}
M.~I. D'yakonov, V.~A. Marushchak, V.~I. Perel', and A.~N. Titkov,
\newblock Soviet Physics - JETP {\bf 63}, 655 (1986).

\bibitem{KhaetskiiPRB00}
A.~V. Khaetskii and Y.~V. Nazarov,
\newblock Phys. Rev. B {\bf 61}, 12639 (2000).

\bibitem{KhaetskiiPRB01}
A.~V. Khaetskii and Y.~V. Nazarov,
\newblock Phys. Rev. B {\bf 64}, 125316 (2001).

\bibitem{WoodsPRB02}
L.~M. Woods, T.~L. Reinecke, and Y.~Lyanda-Geller,
\newblock Phys. Rev. B {\bf 66}, 161318 (2002).

\bibitem{GolovachPRL04}
V.~N. Golovach, A.~Khaetskii, and D.~Loss,
\newblock Physical Review Letters {\bf 93}, 016601 (2004).

\bibitem{BulaevPRB05}
D.~V. Bulaev and D.~Loss,
\newblock Physical Review B {\bf 71}, 205324 (2005).

\bibitem{DresselhausPR55}
G.~Dresselhaus,
\newblock Phys. Rev. {\bf 100}, 580 (1955).

\bibitem{BychkovJoPCSSP84}
Y.~A. Bychkov and E.~I. Rashba,
\newblock Journal of Physics C: Solid State Physics {\bf 17}, 6039 (1984).

\bibitem{Yamamoto99}
Y.~Yamamoto and A.~Imamoglu,
\newblock {\em Mesoscopic Quantum Optics},
\newblock John Wiley \& Sons, 1999.

\bibitem{BrosselJDPELR52}
J.~Brossel, A.~Kastler, and J.~Winter,
\newblock Journal de physique et le Radium {\bf 13}, 668 (1952).

\bibitem{ShabaevPRB03}
A.~Shabaev, A.~L. Efros, D.~Gammon, and I.~A. Merkulov,
\newblock Physical Review B {\bf 68}, 201305 (2003).

\bibitem{Loudon03}
R.~Loudon,
\newblock {\em The Quantum Theory of Light},
\newblock Oxford Science Publications, third edition, 2003.

\bibitem{LuttingerPR56}
J.~M. Luttinger,
\newblock Phys. Rev. {\bf 102}, 1030 (1956).

\bibitem{BesterPRB03}
G.~Bester, S.~Nair, and A.~Zunger,
\newblock Physical Review B {\bf 67}, 161306 (2003).

\bibitem{KarraiSAM03}
K.~Karrai and R.~J. Warburton,
\newblock Superlattices and Microstructures {\bf 33}, 311 (2003).

\bibitem{HogeleDEZ05}
A.~H\"ogele,
\newblock {\em Laser spectroscopy of single charge-tunable quantum dots},
\newblock PhD thesis, LMU Munich, 2005.

\bibitem{AlenAPL06}
B.~Alen, A.~H\"ogele, M.~Kroner, S.~Seidl, K.~Karrai, R.~J. Warburton,
  A.~Badolato, G.~Medeiros-Ribeiro, and P.~M. Petroff,
\newblock Applied Physics Letters {\bf 89}, 123124 (2006).

\bibitem{AlenAPL03}
B.~Alen, F.~Bickel, K.~Karrai, R.~J. Warburton, and P.~M. Petroff,
\newblock Applied Physics Letters {\bf 83}, 2235 (2003).

\bibitem{HogelePRL04}
A.~H\"ogele, S.~Seidl, M.~Kroner, K.~Karrai, R.~J. Warburton, B.~D. Gerardot,
  and P.~M. Petroff,
\newblock Physical Review Letters {\bf 9321}, 7401 (2004).

\bibitem{HogelePE04}
A.~H\"ogele, B.~Alén, F.~Bickel, R.~J. Warburton, P.~M. Petroff, and K.~Karrai,
\newblock Physica E {\bf 21}, 175 (2004).

\bibitem{NewtonAJoP76}
R.~G. Newton,
\newblock American Journal of Physics {\bf 44}, 639 (1976).

\bibitem{DalgarnoAPL06}
P.~A. Dalgarno, J.~McFarlane, B.~D. Gerardot, R.~J. Warburton, K.~Karrai,
  A.~Badolato, and P.~M. Petroff,
\newblock Applied Physics Letters {\bf 89}, 043107 (2006).

\bibitem{LaiPRL06}
C.~W. Lai, P.~Maletinsky, A.~Badolato, and A.~Imamoglu,
\newblock Physical Review Letters {\bf 96}, 167403 (2006).

\bibitem{EblePRB06}
B.~Eble, O.~Krebs, A.~Lemaitre, K.~Kowalik, A.~Kudelski, P.~Voisin,
  B.~Urbaszek, X.~Marie, and T.~Amand,
\newblock Physical Review B {\bf 74}, 081306 (2006).

\bibitem{BulaevPRL05}
D.~V. Bulaev and D.~Loss,
\newblock Physical Review Letters {\bf 95}, 076805 (2005).

\bibitem{WarburtonN00}
R.~J. Warburton, C.~Schaflein, D.~Haft, F.~Bickel, A.~Lorke, K.~Karrai, J.~M.
  Garcia, W.~Schoenfeld, and P.~M. Petroff,
\newblock Nature {\bf 405}, 926 (2000).

\end{thebibliography}

\end{document}